\documentclass[galaxies,article,accept,moreauthors,pdftex]{mdpi} 

\usepackage{upgreek}

\def\ffam {\hbox{$\,.\!\!^{\prime}$}}
\def\ffas {\hbox{$\,.\!\!^{\prime\prime}$}}

\def\ffm {\hbox{$\,.\!\!^{\rm m}$}}

\def \la{\mathrel{\mathchoice   {\vcenter{\offinterlineskip\halign{\hfil
$\displaystyle##$\hfil\cr<\cr\sim\cr}}}
{\vcenter{\offinterlineskip\halign{\hfil$\textstyle##$\hfil\cr
<\cr\sim\cr}}}
{\vcenter{\offinterlineskip\halign{\hfil$\scriptstyle##$\hfil\cr
<\cr\sim\cr}}}
{\vcenter{\offinterlineskip\halign{\hfil$\scriptscriptstyle##$\hfil\cr
<\cr\sim\cr}}}}}
\def \ga{\mathrel{\mathchoice   {\vcenter{\offinterlineskip\halign{\hfil
$\displaystyle##$\hfil\cr>\cr\sim\cr}}}
{\vcenter{\offinterlineskip\halign{\hfil$\textstyle##$\hfil\cr
>\cr\sim\cr}}}
{\vcenter{\offinterlineskip\halign{\hfil$\scriptstyle##$\hfil\cr
>\cr\sim\cr}}}
{\vcenter{\offinterlineskip\halign{\hfil$\scriptscriptstyle##$\hfil\cr
>\cr\sim\cr}}}}}

\firstpage{1} 
\pubvolume{1}
\issuenum{1}
\articlenumber{0}
\pubyear{2021}
\copyrightyear{2021}
\externaleditor{Academic Editor: Michele Bellazzini} 
\datereceived{6 October 2021} 
\dateaccepted{6 December 2021 } 
\datepublished{} 
\hreflink{https://doi.org/} 

\Title{The Interstellar Medium of Dwarf Galaxies}

\TitleCitation{The Interstellar Medium of Dwarf Galaxies}

\Author{Christian Henkel $^{1, 2, 3,}$*\orcidA{}, 
Leslie K. Hunt $^4$\orcidB{} and 
Yuri I. Izotov $^5$\orcidC{}} 

\AuthorNames{Christian Henkel, Leslie K. Hunt, Yuri I, Izotov}

\AuthorCitation{Henkel, C.; Hunt, L.K.; Izotov, Y.I.}

\address{%
$^{1}$ \quad Max-Planck-Institut f{\"u}r Radioastronomie, Auf dem H{\"u}gel 69, 53125-Bonn, Germany \\
$^{2}$ \quad Department of Astronomy, Faculty of Science, King Abdulaziz University, P.O. Box 80203,  Jeddah 21589, Saudi Arabia \\
$^{3}$ \quad Xinjiang Astronomical Observatory, Chinese Academy of Sciences, Urumqi 830011,  China \\
$^{4}$ \quad INAF-Osservatorio Astrofisico di Arcetri, Largo E. Fermi, 5, 50125 Firenze, Italy; leslie.hunt@inaf.it \\
$^{5}$ \quad Bogolyubov Institute for Theoretical Physics, National Academy of Sciences of Ukraine, 14-b Metrolohichna Str., UA-03143 Kyiv, Ukraine; yizotov@bitp.kiev.ua 
 }

\corres{Correspondence: chenkel@mpifr-bonn.mpg.de}

\abstract{Dwarf galaxies are by far the most numerous galaxies in the Universe, showing properties that are quite different from those 
of their larger and more luminous cousins. This review focuses on the physical and chemical properties of the interstellar medium of those dwarfs 
that are known to host significant amounts of gas and dust. The neutral and ionized gas components and the impact of the dust will be
discussed, as well as first indications for the existence of active nuclei in these sources. Cosmological implications are also addressed, 
considering the primordial helium abundance and the similarity of local Green Pea galaxies with young, sometimes protogalactic sources 
in the early Universe.} 


\begin{document}
\setcounter{section}{0} 
\end{paracol}
%

\section{Introduction} \label{S1}

Luminous massive stars are rare and experience a remarkable but short lifetime. Low mass stars, even though occasionally flaring, are much less 
conspicuous but are numerous, and have lifetimes that usually exceed a Hubble time; they are also able to conserve much of their original composition. 
Galaxies are, in some respect, qualitatively similar. The most massive ones are also rare and luminous. After a relatively brief active phase they 
tend to become quiescent giant ellipticals, `cD' or `D' galaxies, exhibiting (if at all) only little star formation during the current epoch. Small 
galaxies, like their low mass stellar cousins, dominate not only in number (e.g., \cite{1,2,3}) but can also retain in many cases significant amounts of their 
original elemental composition, eventually even re-activating larger galaxies through minor mergers. In exceptional cases they can stay isolated up 
to the present day. Dwarf galaxies are therefore of high importance not only in their own right: because of their physical, chemical and 
kinematical properties they also serve as proxies of the distant, early and chemically-unevolved Universe. Furthermore, their spatial distribution 
helps to constrain models of the evolution of the Universe. Thus dwarf galaxies are not only relevant for a better understanding of our `neighborhood', 
addressing ongoing local physical, chemical and dynamical processes, but they also have the potential to shed light onto the distant past with otherwise 
inaccessible linear resolution.  

The generally accepted rule to define the term `dwarf galaxy' is based on luminosity rather than size. Clearly, dwarfs must be small and dim with respect to Milky 
Way sized major galaxies and bright and extended with respect to old globular clusters like those encountered in the Galaxy. Small galaxies are often irregularly 
shaped and sometimes severely stretched by interactions. Therefore luminosity, not size, is commonly chosen as a criterion to identify dwarf galaxies. In case of a low 
level of obscuration by dust (see below), this is also an indicator of mass even though mass-to-luminosity ratios may vary substantially in view of the presence
or absence of young massive stars (e.g., \cite{4}). In the following we consider an upper luminosity bound of $M_{\rm V}$ $\approx$ $-$18$^{\rm M}$ ($L_{\rm V}$ 
$\approx$ 1.4 $\times$ 10$^9$\,L$_{\rm V,\odot}$), unless otherwise noted. This implies that the bulk of galaxies being able to form spiral disks are not included. 
Galaxies like the Large Magellanic Cloud (LMC) and the Triangulum galaxy (M\,33) in the local and M\,82 in in the nearby ($D$ $\approx$ 3.5\,Mpc) M\,81 group are 
slightly too luminous. However, the Small Magellanic Cloud (SMC) lies well below the above mentioned upper luminosity threshold. At the lower end, we assume that 
a dwarf galaxy with a well developed stellar component must be significantly more luminous than a typical globular cluster in the Milky Way. This implies $M_{\rm V}$ 
$\ll$ $-$8$^{\rm M}$ (for dwarfs with an even lower stellar luminosity, being almost always devoid of a notable interstellar medium, \mbox{see \cite{5}}). Massive 
objects mainly consisting of neutral hydrogen and lacking a significant stellar component appear to be statistically irrelevant (\cite{6}; see also Section \ref{S4.3}), 
while (damped) Ly-$\alpha$ systems are beyond the scope of this paper.

In Section \ref{S2}, we describe the main classes of dwarf galaxies and their characteristics. The main components of their interstellar medium are introduced in Section \ref{S3}. 
An overview of the interplay of these different components, modes of star formation and the presence of active galactic nuclei are the topic of Section \ref{S4}. Section \ref{S5}
introduces individual star-forming dwarfs in the local Universe, while some cosmological aspects are discussed in Section \ref{S6}. Finally, attractive future prospects and 
promising avenues for future research are summarized in Section \ref{S7}.



\section{Dwarf Galaxies: General Properties} \label{S2}

\subsection{Morphological Types} \label{S2.1}

Prior to a detailed analysis of the interstellar medium (ISM) of dwarf galaxies, we first have to address their heterogeneity. Dwarf ellipticals 
(dEs), a prominent group among early-type galaxies, populate like their more massive cousins a `fundamental plane' in three dimensional 
space, defined by the central projected velocity dispersion, a linear scale (e.g., the `core', `half-light' or `effective' radius) and a 
surface brightness (averaged over the core, half-light or effective radius). Nevertheless they are quite distinct from giant ellipticals, 
since they become more diffuse with decreasing luminosity and do not follow the de Vaucouleurs $r^{1/4}$  but more closely an exponential 
light distribution. Furthermore, these smaller galaxies do not occupy the same fundamental plane and show a larger scatter in their parameters
(e.g., \cite{7,8}) due to variations in the mass-to-luminosity ratios and structural parameters. The morphological transition from E to dE is near 
$M_{\rm B}$  = $-$18$^{\rm M}$, but there is overlap. M32, the compact satellite of the Andromeda nebula M31, still follows the de Vaucouleurs 
law (e.g., \cite{9}). Some dS0 galaxies also exist, which, below, will be lumped together with the dE objects. Many of the more luminous dEs ($M_{\rm B}$ 
$\la$ $-$16$^{\rm M}$) are `nucleated', possessing a central star cluster sometimes reaching $M_{\rm V}$ = $-$12$^{\rm M}$ (see e.g., NGC\,205 in 
Section \ref{S5.2}). Some of these galaxies also show evidence for the existence of disks and spiral structure and bars {(e.g., \cite{10,11,12})}.
While some of the dEs appear to be rotationally supported, others appear to be pressure supported and there are cases with kinematically 
decoupled cores (e.g., \cite{13,14}).  

In our Local Group, dwarf spheroidal galaxies (dSphs) are numerous. Being much more extended than globular clusters, they are faint, difficult 
to find, commonly lack a detectable interstellar medium and are therefore not a major topic of this review. Typically, they did not form 
stars since several hundred million years or more (e.g., \cite{15}). A lack of a notable interstellar medium also characterizes the even less luminous 
ultra-diffuse galaxies (UDGs) and ultra-faint dwarfs (UFDs) at $M_{\rm V}$ $\ga$ $-$7.7$^{\rm M}$ ($L$ $\la$ 10$^5$\,L$_{\odot}$; \cite{5}) as well 
as the ultra-compact dwarfs, the latter with structural properties that resemble those of $\omega$\,Cen (e.g., \cite{5,15,16}).

There are not only `early-type' but also `late-type' galaxies of low luminosity, often following an exponential surface brightness
profile. These are commonly termed dwarf irregular galaxies (dIrrs) with the Small Magellanic Cloud (SMC) and NGC\,6822 as the nearest such objects. 
Some of them, like the SMC, owe their shape to a recent interaction with another galaxy. The majority of this class of objects is characterized by 
a massive interstellar medium that is dominated by neutral hydrogen (H{\sc i}; {see e.g., \cite{17}}). Among this class of late-type objects, 
blue compact dwarf galaxies (BCDs) stand out because they host sites of recent or ongoing massive star formation\footnote{e.g., 
ned.ipac.caltech.edu/level5/Sept02/Palco-BCD/Agpaz3.html}, by a high surface brightness, e.g., $\mu_{\rm B,peak}$ $<$ 22$^{\rm m}$/arcsec$^2$, 
and blue color, e.g., $\mu_{\rm B,peak} - \mu_{\rm R,peak}$ $\la$ 1$^{\rm m}$. Even more active but sometimes showing luminosities exceeding 
our definition of a dwarf galaxy (\mbox{Section \ref{S1}}) are the `Green Peas' (also termed Luminous Compact Galaxies, LCGs), compact metal 
poor galaxies of size $\approx$5\,kpc or less, that are vigorously forming stars (e.g., \cite{18}). Their color is related to intense emission 
in the [O{\sc iii}]$\lambda$5007 line and is particularly dominant at optical wavelengths, when redshifts move the H$\alpha$ line out of the optical 
spectral window. Naturally, the green color refers to those galaxies detected at moderate redshifts ($z$ $\approx$ 0.2) and changes at lower and 
higher redshifts. 

It is sometimes difficult to distinguish quiescent dIrr galaxies from their dE counterparts. Furthermore, there exist real transitory objects, 
like the Phoenix system (e.g., \cite{9}). Early-type dwarfs can be former late-type systems that have lost their gas in a crowded environment (e.g., 
\cite{19,20}). While the Local Group contains several dIrr galaxies, BCDs are rarer and only a single such object, IC\,10, has been suggested to be 
located within its volume \cite{21}. Green peas are even rarer than BCDs and are therefore commonly observed at redshifts $z$ $>$ 0.1. Not 
surprisingly, dIrrs preferentially reside in field environments, while dEs are more commonly found in denser more clustered regions (e.g., 
\cite{15,17}; for the Virgo cluster, see e.g., \cite{22}). The latter also holds for the relatively luminous nucleated dwarfs.

\subsection{Physical and Chemical Boundary Conditions} \label{S2.2}

As a start, we have to mention basic differences between the ISM of dwarfs and giant galaxies.
For dwarfs:
\begin{itemize}
\item The ISM is metal poor.
\item Gravity is weak.
\item Interstellar pressure is low.
\item As a consequence, disks tend to be thick and diffuse.
\item Low gravity and pressure may lead to strong feedback effects.
\item There is little shear due to rotational effects.
\item There is no high-contrast spiral structure shocking and compressing the gas.
\item Dust to gas mass ratios ($M_{\rm d}$/$M_{\rm g}$) are low.
\item Insufficient dust shielding leads to a harsh environment for molecular species.
\item Low $M_{\rm d}$/$M_{\rm g}$ ratios lead to a low gas phase depletion of refractory elements onto dust grains. 
\end{itemize} 

\begin{figure}[H]	
\includegraphics[width=13.5cm]{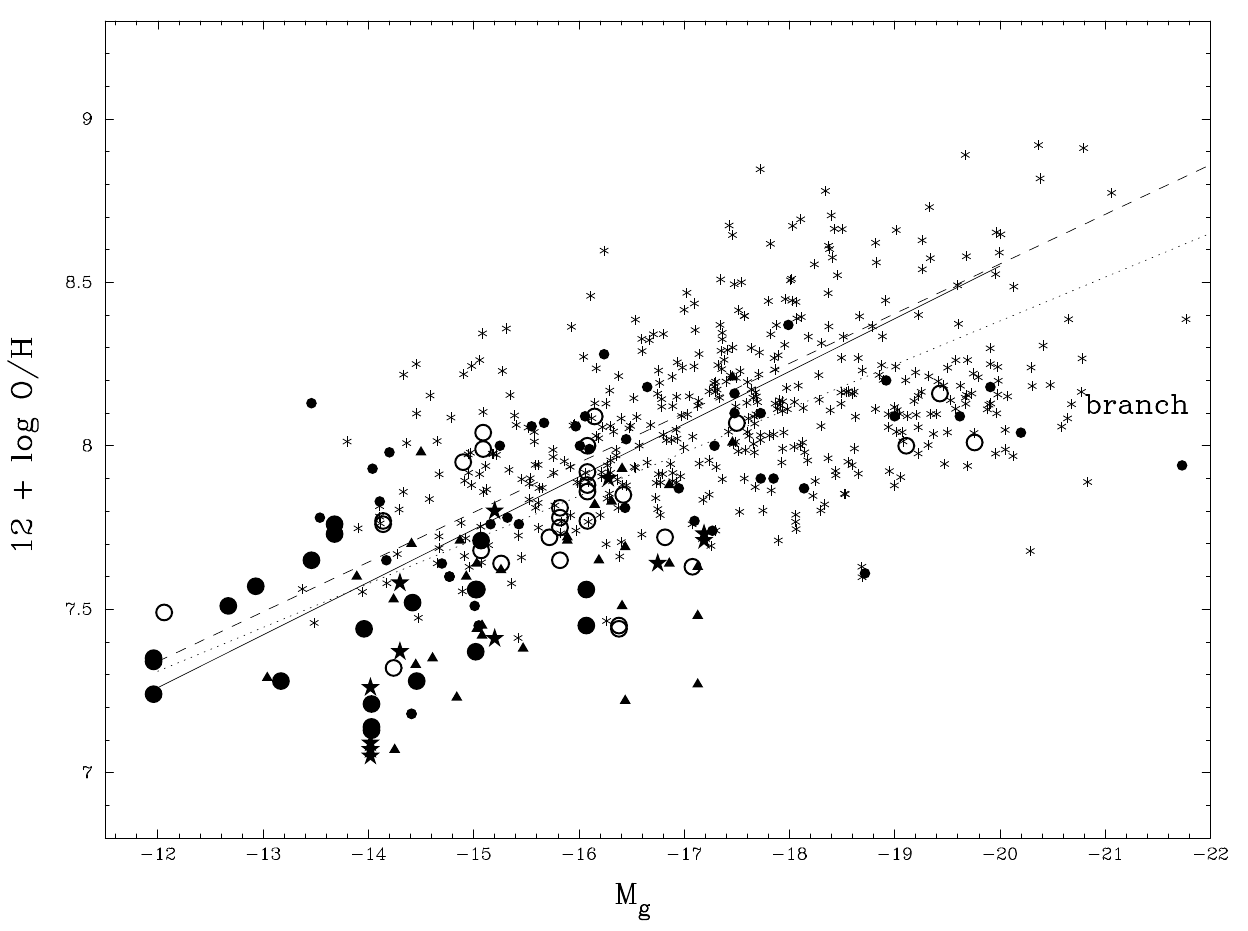}
\caption{Oxygen abundances of mostly local galaxies as a function of absolute g-magnitude. `Branch' on the right hand side indicates particularly 
distant objects ($z$ $>$ 0.2). Small filled circles: data from \cite{33,34}; large filled circles: data from \cite{35}; asterisks: a Sloan Digital 
Sky Survey (SDSS) sample; open circles: 3.6\,m ESO data; stars: VLT data; triangles: additional VLT data from \cite{36}. For more details, see \cite{23}.}
\label{fig1}  
\end{figure}

\smallskip
For mass-metallicity and luminosity-metallicity correlations (Figure~\ref{fig1}), metallicities are mainly deduced from optical spectra of H{\sc ii} regions, 
see e.g., \cite{23,24,25,26}. Because of their small masses, rotation speeds in dwarfs are low, far below $v_{\rm rot}$ = 100\,km\,s$^{-1}$, which implies 
that random dynamical events can play a significant role in shaping the detailed velocity field. Late-type dwarf irregulars tend to show rotation, even 
though sometimes with very low rates ($v_{\rm rot}$/$\sigma$ $<$ 1; $\sigma$: velocity dispersion; e.g., \cite{27})). On the other hand, it is hard to 
find rotating dwarf spheroidals. This also holds for the larger dwarf ellipticals NGC\,147, NGC\,185, and NGC\,205 that accompany the Andromeda nebula 
(e.g., \cite{15,28}). A deeper consideration of rotational properties would require a discussion of dark matter and alternative gravitational theories, 
which is the topic of accompanying articles and therefore beyond the scope of this work. This also holds for a detailed discussion of the Tully-Fisher 
relation. Here we focus instead on directly observable ISM properties, starting with the most extended and also by mass most dominant interstellar component, 
the atomic hydrogen (H{\sc i}).  This is followed by discussions of the the dust and molecular gas components to then also address relevant properties 
of the ionized medium. Not mentioned below but also relevant for this study is that we assume a solar system metallicity of $Z$ = 12 + log(O/H) = 8.7 to 
8.9 \cite{29} and that in small galaxies showing little nuclear processing carbon and nitrogen tend to be even more depleted than oxygen (e.g., \cite{30,31,32}).

\section{Main Components of the Interstellar Medium} \label{S3}

\subsection{Neutral Hydrogen (H{\sc i})} \label{S3.1}

Several large surveys of the $\lambda$ $\approx$ 21\,cm line, with arcsecond resolution and channel widths of only a few km\,s$^{-1}$, have been 
carried out to unveil the properties of H{\sc i} in galaxies of low luminosity. Over the last 10 to 15 years, the most important of these include 
FIGGS (Faint Irregular Galaxies GMRT Survey; \cite{37}) with the Giant Meter Radio Telescope covering $\approx$60 nearby ($D$ $\la$ 4\,Mpc) dwarfs 
and the {\it JVLA} ({\it Jansky Very Large Array}) projects THINGS (The H{\sc i} Nearby Galaxy Survey; \cite{38}) with 12 dwarfs among the 34 targeted sources, 
SHIELD (Survey of H{\sc i} in Extremely Low mass Dwarfs; \cite{39}) with 12 sources, and the {\it JVLA}-ANGST survey (Very Large Array survey of ACS Nearby 
Galaxy Survey Treasury galaxies; \cite{40}) encompassing 35 dwarfs at $D$ $\la$ 4\,Mpc. The Little Things survey (Local Irregulars That Trace Luminosity 
Extremes; \cite{41}), again obtained with the {\it JVLA}, presents observations of 37 dwarf irregulars and 4 BCDs, while a compilation of local volume objects 
(distance $D$ $<$ 11\,Mpc), comprising a total of 1072 galaxies, among them many dwarfs, has been presented by \cite{17}.  

\begin{figure}[H]	
\includegraphics[width=13.5cm]{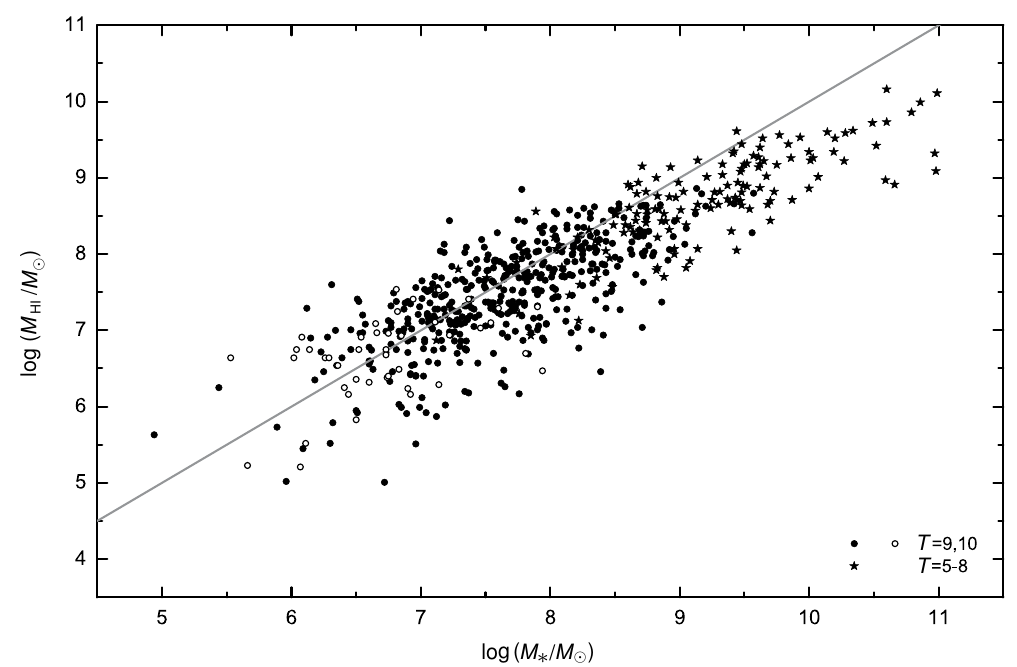}
\caption{Neutral hydrogen versus stellar masses of late-type galaxies ($T$ = 5--10) in the local volume ($D$ $<$ 11\,Mpc). Open circles denote 
galaxies with upper H{\sc i} flux limits. The solid line connects points with equal H{\sc i} and stellar masses from \cite{17}.} 
\label{fig2}
\end{figure}

Among these 1072 nearby galaxies, 52\% are late-type and 34\% early-type dwarfs, amounting to more than 80\% of the sample. Half of these are
located near major galaxies like the Milky Way or M\,81. While spheroidal dwarfs show a tendency to be located in more crowded environments, 
there also exist rare exceptions from this rule with dSphs quite far from any major source, but being nevertheless devoid of a significant 
interstellar medium \cite{17}. H{\sc i} is detected in 91\% (424/467) of the H{\sc i} observed late-type but only in 7\% (13/180) of the 
H{\sc i} observed early-type dwarfs. Plotting exclusively late type galaxies ($T$ = 5--10, following the revised Hubble system), the 
H{\sc i} to stellar mass ratio ($M$(H{\sc i})/$M_*$; see Figure~\ref{fig2}) is to first order constant and close to unity. While not unexpectedly 
decreasing to values well below unity in massive galaxies, a large scatter would also be introduced if values far below unity would be included 
from dSph and dE galaxies at the low mass end.

{\it JVLA} observations of the BCD Haro\,11 \cite{42} showed an intriguing result where no emission but rather absorption has been detected towards the 
center of the continuum emission including the location of the so-called knot B. Possibly, this is the first dwarf galaxy showing H{\sc i} 
absorption against its own central continuum emission, reminiscent of a situation commonly encountered in radio galaxies (e.g., \cite{43}). 
To discuss larger samples, \cite{44} measured with the Green Bank 100 m telescope 29 galaxies with 12 + log(O/H) $\leq$ 7.6 and obtained detections 
in almost all of them (28/29). They find an increasing gas mass fraction with decreasing luminosity, mass and metallicity, dominated by H{\sc i}. 
\cite{45} analyzed within the framework of the Little Things project the H{\sc i} emission of three seemingly isolated BCDs to find out what is 
triggering their enhanced star formation activity. An obvious interaction with other sources is not found, but advanced mergers, strong 
stellar feedback and/or ram pressure stripping might be viable possibilities. 

A first H{\sc i} survey of Green Pea galaxies at redshifts $<$0.1 was recently published \mbox{by \cite{46}}. For the detected 19 sources (a 50\% detection 
rate) the $M$(H{\sc i})/$M_*$ mass ratios are quite high and consistent with expectations for such active compact dwarf galaxies (see Figure~\ref{fig2}). 
However, the atomic gas depletion timescale is extremely short, with $\tau_{\rm dep}$ = $M$(H{\sc i})/SFR $\approx$ 0.6\,Gyr (SFR denotes the 
star-forming rate). This is only about 10\% of the commonly attained values by main sequence galaxies showing standard specific star formation
rates (sSFR = SFR/$M_*$).

\subsection{The Dust Component}  \label{S3.2}

The {\it InfraRed Astronomical Satellite} ({\it IRAS}) with wavelengths of 12.5--100\, $\upmu$m, the {\it Infrared Space Observatory} ({\it ISO}) at 
2.5--240\, $\upmu$m, and the {\it Spitzer} mission (3.6--180\, $\upmu$m) provided fundamental insights into the extragalactic infrared (IR) sky, revealing 
galaxies that are emitting electromagnetic radiation almost exclusively in the infrared. However, while these observatories had apertures of 57, 60 
and 85\,cm, respectively, it is the {\it Herschel Space Observatory} with its superior 3.5\,m mirror that provided more recently, from 2009 to 2013, 
the deepest infrared views into space with a wavelength range of 70--500\, $\upmu$m. While dust is, even compared with the gas, only a tiny component by mass, 
it dominates the infrared emission and is an essential ingredient to soften radiation emitted at higher frequencies through absorption and re-emission, 
to protect molecules from photodissociation (\mbox{Section \ref{S3.4}}), and to provide appropriate cooling for an ISM mainly heated by stellar activity.  

Dust grains require metals to form (e.g., \cite{47}). These are synthesized in stars, being released into the ISM either on short time scales 
by supernovae or on longer time scales during the asymptotic giant branch phase of stars with lower mass ($M$ $<$ 8\,M$_\odot$). Dust is accumulated 
during quiescent phases in well shielded regions and processed in harsher environments, including energetic stellar radiation, shocks and stellar or 
galactic winds. At very low metallicities of 12 + log(O/H) $<$ 8.0, dust grains are difficult to form due to the harsh conditions in an almost unshielded 
ISM. Only at higher metallicities the fractional interstellar dust grain abundance can grow approximately proportionally to the fractional oxygen abundance 
(e.g., \cite{48,49}). At lower metallicities, the dust-to-gas mass ratio tends to decrease faster than the oxygen abundance (see also the model
calculations of \cite{50}) but there may be exceptions \cite{51,52}.  

It was early recognized that dwarf galaxies host hotter dust but less emission from so-called polyaromatic hydrocarbons (PAHs) with their numerous bands 
in the mid-infrared, the most prominent of them located at 6.2, 7.7, 8.6, 11.3 and 12.7  $\upmu$m. There is a gradual decrease in PAH emission with decreasing 
metallicity, while the far-infrared color temperature of the large dust grains reaches a peak near 12 + log(O/H) = 8.0 (e.g., \cite{48,53,54}). 50 local dwarf 
galaxies were studied by the {\it Herschel} Dwarf Galaxy Survey and compared with dust properties of larger galaxies \cite{55,56,57}. A characteristic long 
wavelength ($\lambda$ $>$ 20   $\upmu$m) emissivity-index of $\beta_{\rm IR}$ = $-$1.7 was obtained, that does not strongly depend on metallicity ($\beta_{\rm IR}$ 
is defined by $S$ = $\nu^{\beta_{\rm IR}}$ $\times$ $B$($\nu$,$T_{\rm dust}$) with $S$ denoting the flux density, $\nu$ the frequency, and $B$ representing 
the Planck function). Because of higher temperatures, for dwarfs the dust emission per mass unit can be several times stronger in the FIR/submm bands than 
in galaxies of approximately solar metallicity. The dust morphology is clumpy, and in dwarfs mainly heated by massive stars (e.g., \cite{47}). In cases 
of high H$\beta$ equivalent widths and luminosities, there can be dust components that reach, based on {\it WISE (Wide-field Infrared Survey Explorer)} 
temperatures of several 100\,K (\cite{58}; for an individual source, see also \cite{59}). 

Dust obscures emission at wavelengths shorter than the size of the grains, thus can affect fundamental parameters obtained at near infrared, 
optical and UV wavelengths. However, the effect appears not to be large in most sources. Internal extinction is apparently insignificant in most 
dwarf galaxies (e.g., \cite{32}), while \cite{58} find a total (Galactic extinction plus the intrinsic extinction of the extragalactic object) of $A_{\rm V}$ = 
0$^{\rm m}$\!\!\!.\,6 as a typical value for their large sample of $\approx$14,000 SDSS (Sloan Digital Sky Survey) galaxies with an average oxygen 
abundance of 12 + log(O/H) = 7.95. Although optical recombination lines tend to give fairly low values of $A_{\rm V}$ in dwarf galaxies, near- and mid-
infrared recombination lines give much higher values (e.g., \cite{32,60,61}) implying that at least some dwarfs host dust-embedded H{\sc ii} regions despite 
their low metallicity.

\begin{figure}[H]	
\includegraphics[width=13.5cm]{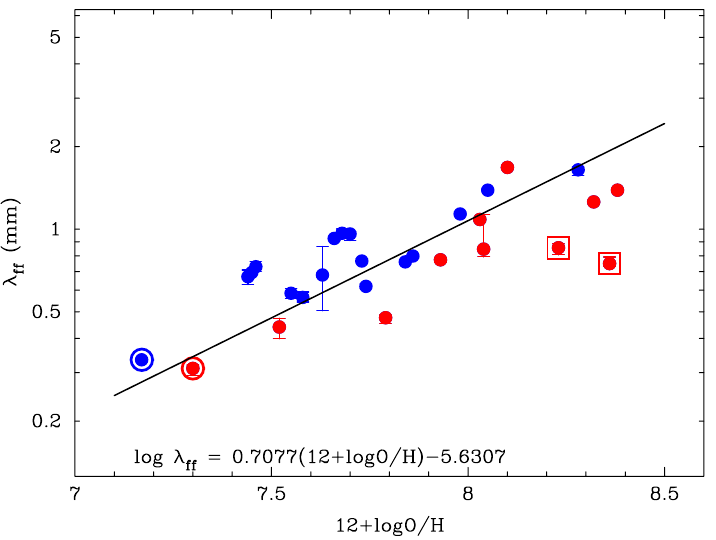}
\caption{The wavelength at which dust and free-free continuum emission reach the same intensity level is plotted as a function of metallicity for 
galaxies with $L$(H$\beta$) $<$ 10$^7$\,L$_{\odot}$ (blue filled circles) and $L$(H$\beta$) $>$ 10$^7$\,L$_{\odot}$ (red filled circles). The 
correlation demonstrates how strongly metallicity affects the dust to free-free emission ratios, while the specific star formation activity of 
individual galaxies is leading to significant scatter. The encircled galaxies in the lower left are IZw\,18 (blue) and SBS\,0335--052 (red). The 
vigorously star forming galaxies Haro\,11 and IIZw\,40 are indicated by surrounding squares. The solid line is an unweighted fit to the shown 
{\it Herschel} sample with a total of 28 galaxies \cite{62}.}
\label{fig3}
\end{figure}

\subsection{The Submillimeter Excess}  \label{S3.3}

At $\ga$ 500 $\upmu$m, continuum levels are obtained that surpasss expected flux densities deduced from the measured dust emission obtained at shorter 
wavelengths by significant $>$10\%. To give an example: 41\% of the 22 targets of the {\it Herschel} Dwarf Galaxy Survey detected at 500 $\upmu$m showed 
the excess \cite{56}, when making use of the \cite{63} models of dust emission. This implied that the effect is not common to all galaxies but becomes more 
pronounced in sources of lower metallicity. A large number of potential explanations were put forward (e.g., \cite{47,56,64,65,66,67}): 
\begin{itemize}
\item A particularly cold so far missed dust component.
\item A long wavelength enhancement of the opacity of silicate grains.
\item A kinetic temperature dependent grain emissivity, with the emissivity index $\beta_{\rm IR}$ decreasing when $T_{\rm kin}$ increases.
\item Spinning grains.
\item Cosmic microwave background (CMB) fluctuations.
\item A positive correlation with the H{\sc i} component.
\item Blending with CO lines.
\end{itemize}

While a detailed discussion of all these suggestions is well beyond the scope of this paper, it is worth noting that the introduction of a so far 
missed very cold dust component would lead to very high $M_{\rm dust}$/$M_{gas}$ ratios, because at low temperatures dust emissivity per dust mass 
is very low. This cannot be reconciled with the fact that in metal poor galaxies $M_{\rm dust}$/$M_{\rm gas}$ values should be particularly low 
because dust grains require the inclusion of heavy elements (e.g., \cite{47}). A similar argument also holds for CO (see Section \ref{S3.4}).

Based on measurements of the H$\beta$ line, a solution to this puzzle has been presented by \cite{62}. After removing the stellar absorption 
component and accounting for extinction, the H$\beta$ line permits the derivation of massive star formation rates, which also determines the flux 
densities of the free-free emission. This so-called bremsstrahlung can be significant in galaxies of low metallicity: because of high temperatures 
and low $M_{\rm dust}$/$M_{\rm gas}$ ratios, dust emission may start to dominate the continuum emission only at shorter wavelengths and higher 
frequencies than in larger more metal rich galaxies. Prior \mbox{to \cite{62}} this was already mentioned by \cite{68,69}. However, lacking accompanying 
H$\beta$ data, they concluded that free-free emission should be insufficient to explain the submillimeter excess. The more metal poor a galaxy is, 
inhibiting a strong dust continuum at submillimeter wavelengths, and the more active the galaxy is with respect to star formation, enhancing the 
free-free emission, the more dominant bremsstrahlung becomes, creating a detectable `excess' of continuum emission at wavelengths of $\ga$500 $\upmu$m 
(Figure~\ref{fig3}; see also \cite{70} and Section \ref{S5.2}).

\subsection{The Molecular Component} \label{S3.4}

Showing (unlike H$_2$) a small but notable dipole moment, being heavy enough to reveal rotational transitions at mm- and submm-wavelengths and belonging 
to one of the most abundant species, CO is the most common tracer of molecular gas, in the Milky Way as well as in extragalactic objects. Thus it traces 
best the material being used for future star formation. The first CO map obtained from a dwarf galaxy, covering 2$^{\circ}$ $\times$ 2$^{\circ}$ and 
published by \cite{71} with a resolution of 8\ffam8, revealed the overall distribution of the molecular gas in the SMC. It indicated a low mass ratio 
($\approx$7\%) of molecular (predominantly H$_2$) to atomic (predominantly H{\sc i}) neutral gas and a high H$_2$ column density to integrated CO 
intensity conversion factor of $X$(CO) = $N$(H$_2$)/$I$(CO) = 6 $\times$ 10$^{21}$\,cm$^{-2}$\,K$^{-1}$\,km$^{-1}$\,s, $\approx$30 times the Galactic 
disk value (\cite{72}). Nowadays, much higher sensitivities and angular resolutions can be achieved, basically confirming the trend suggested by this 
early paper (see below). 

\begin{figure}[H]	
\includegraphics[width=13.5cm]{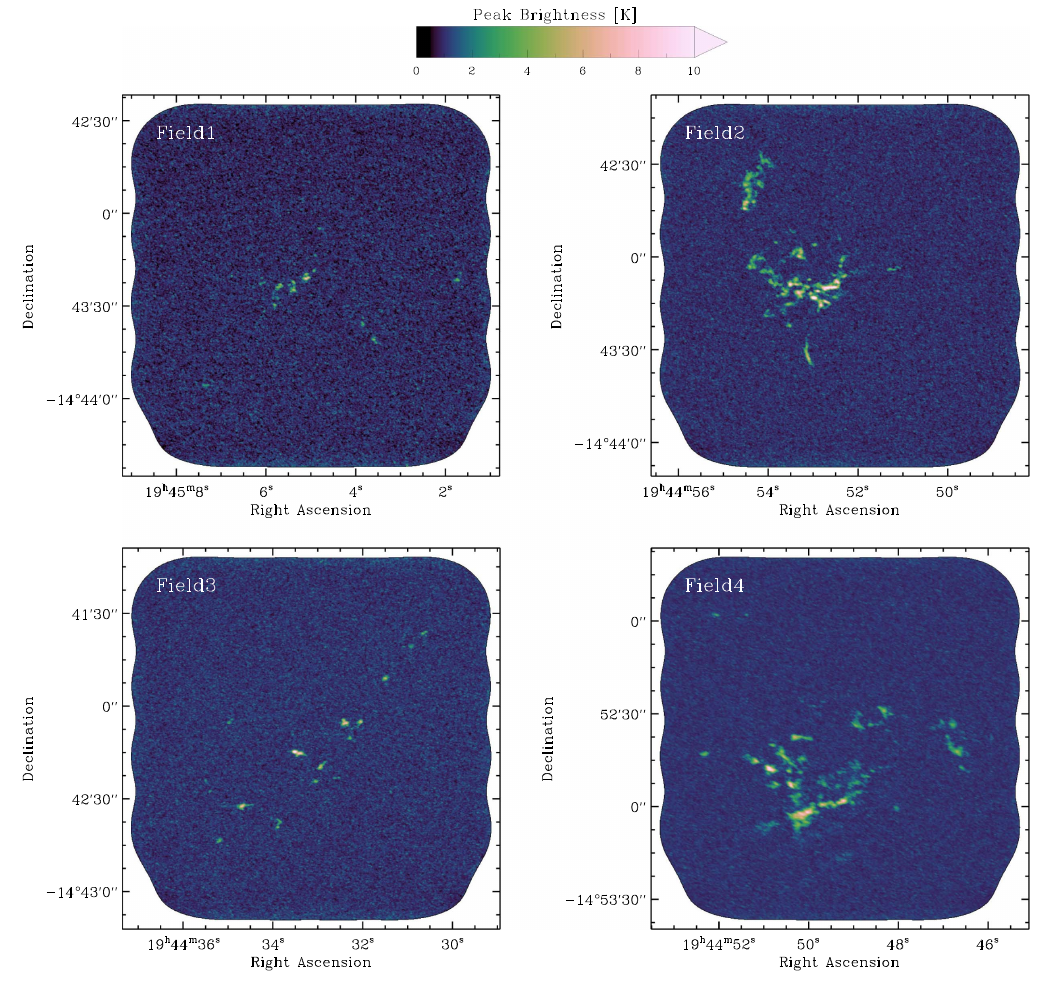}
\caption{{\it ALMA} data with 2\,pc linear resolution (0\ffas9) of four selected regions with star-forming activity inside of NGC\,6822 
taken from \cite{77}.} 
\label{fig4}
\end{figure}

The lower the metallicity becomes, the more difficult it is to detect the molecular component of the gas. In galaxies with fractional oxygen abundances 
below that of the SMC (12 + log(O/H) $\approx$ 8.2), very little CO is detected even in galaxies showing vigorous star formation (see, e.g., the patchy
morphology of CO in NGC\,6822 shown in \mbox{Figure~\ref{fig4}}). The problem is that the shielding properties of CO are far weaker than those of H$_2$ 
because of much lower column densities. The lower the metallicity, the less carbon and oxygen are available to form CO and the lesser amounts of dust are 
available to shield these fragile molecules against the interstellar radiation field. When measured with single-dish beam sizes, i.e., with the {\it IRAM} 
30\,m, this leads to $\alpha$(CO) = $M_{\rm H2}$/$L$(CO) $\propto$ ($Z$/Z$_\odot$)$^{-{\rm y}}$ with \mbox{y $\approx$ 3.3} \cite{48}, $\approx$ 1.5 
\cite{73} or $\approx$ 2.0 \cite{74}. As a consequence, there must exist large amounts of CO-dark molecular gas. Such gas is already known to cover 
significant volumes of molecular clouds in the solar neighborhood with its relatively high metallicity \cite{75}. It should become even more predominant 
in dwarf galaxies with gas lacking carbon, oxygen and other metals. Therefore, in spite of often optically thick mm-wave CO lines, where a lower column 
density would not drastically change the peak line intensity, the volume of the region from where CO is emitted decreases and thus leads overall to 
reduced line intensities as long as the emitting region is not spatially resolved \cite{72}. X(CO) and $\alpha$(CO) thus also depend on the angular 
resolution \cite{76,77}. 

To directly relate H$_2$ mass to CO column density independently of a conversion factor, the virial theorem can be applied to individual clouds
as long as they are spatially resolved. However, its application should be viewed as an estimate of the mass confined to the CO emitting volume 
and not revealing this critical value for the entire molecular cloud or complex (e.g., \cite{72}). A better way to quantify the total molecular gas mass 
is to combine several interstellar tracers (see Section \ref{S4.2}).

As mentioned, CO observations become more and more difficult when the metallicity decreases. An interesting set of galaxies is that of \cite{74}, 
who detected CO in eight dwarf galaxies outside the Local Group that cover a metallicity range of 12 + log(O/H) = 7.7 -- 8.4. Nevertheless, 
the signal-to-noise ratio of the CO $J$ = 1$\rightarrow$0 line of the lowest metallicity galaxy in this sample, CGCG\,007--025, is with $\approx$3 
quite low and indicates that at such a metallicity, $\la$10\% of the solar one, the {\it IRAM} 30\,m is close to its sensitivity limit. For a 
similarly sized sample of BCDs (although at slightly higher metallicities) also studied with the {\it IRAM} 30\,m, see \cite{73}.

To date, {\it ALMA}, the {\it Atacama Large Millimeter/submillimeter Array}, and {\it NOEMA}, the {\it Northern extended Millimeter Array}, 
provide by far the highest accessible sensitivity and angular resolution. A few exemplary dwarf galaxies have already been studied. The nearest 
one, still belonging to the Local Group, is NGC\,6822 ($D$ $\approx$ 474 $\pm$ 13\,kpc; \cite{78}). With a sub-SMC metallicity of 12 + log(O/H) 
= 8.02 $\pm$ 0.05, NGC\,6822 has been observed in the CO $J$ = 2$\rightarrow$1 transition \cite{77} with an angular resolution of 0\ffas9, 
corresponding to unprecedented 2\,pc (Figure~\ref{fig4}) beyond the Magellanic Clouds. Observing four 250\,pc sized areas that amount to about two thirds 
of the star formation activity in this galaxy reveal $\approx$150 compact CO clumps with sizes of only 2--3\,pc and full half power line widths of 
$\approx$1\,km\,s$^{-1}$. The CO clumps only cover a tiny part of the observed area and there are no such clumps in a fifth field outside the 
active zone of NGC\,6822. Apparently, CO can only survive in tiny particularly well shielded pockets of interstellar gas. The clumps themselves 
follow with respect to size, column and spatial density and dynamics similarly sized clumps in the Milky Way. Only when considering larger linear 
scales does the low metallicity start to play a dominant role. An even more extreme object, the irregular galaxy Wolf-Lundmark-Melotte (WLM) with 
12 + log(O/H) $\approx$ 7.8, studied by \cite{79}, revealed 10 isolated CO $J$ = 1$\rightarrow$0 clumps with properties similar to those encountered 
in NGC\,6822. One of the most extreme CO detected galaxies with respect to metallicity (12 + log(O/H) = 7.5; \cite{80}), Sextans~B (also known as DDO\,70), 
has been mapped with {\it ALMA} by \cite{81}. Using a linear resolution of 1.4\,pc for this source located in the outskirts of the Local Group revealed 
five clumps, accounting for the bulk of the single-dish emission and following Larson's size-line width relation for the disk of the Milky Way.

The prototypical BCD IIZw\,40 with 12 + log(O/H) = 8.1 and $D$ $\approx$ 10\,Mpc has been studied by \cite{82}, using {\it ALMA} in the CO $J$ = 
1$\rightarrow$0 to 3$\rightarrow$2 lines. The system appears to have undergone a major merger. Beam sizes of 0\ffas5 correspond to linear 
resolutions of 24\,pc, channel widths were 2\,km\,s$^{-1}$. While there is a number of spatially unresolved CO clumps, the molecular emission 
is dominated by a central almost 100\,pc sized elongated structure exhibiting line intensity ratios that favor the $J$ = 3$\rightarrow$2 line 
and that are more reminiscent of Luminous InfraRed Galaxies (LIRGs) than normal star-forming objects. Possibly due to the interaction, the
size-linewidth relation deviates from that in the disk of the Milky Way. Finally, Mrk\,71, also known as NGC\,2363  and located at a distance 
of 3.4\,Mpc \cite{83}, contains two Super Star Clusters (SSCs). The clumpy CO condensations observed include a 
component, exhibiting two velocity features, that coincide in projection with the SSC Mrk\,71-A and may undergo momentum-driven feedback. 
For more information on SSCs, the present day counterparts of forming globular clusters, and their relation to a dense metal poor ISM, 
see the paragraphs on NGC\,1140, SBS0335--052, and NGC\,5253 in \mbox{Section \ref{S5.2}.}

\subsection{The Ionized Gas} \label{S3.5}

\begin{figure}[H]	
\includegraphics[width=13.4cm]{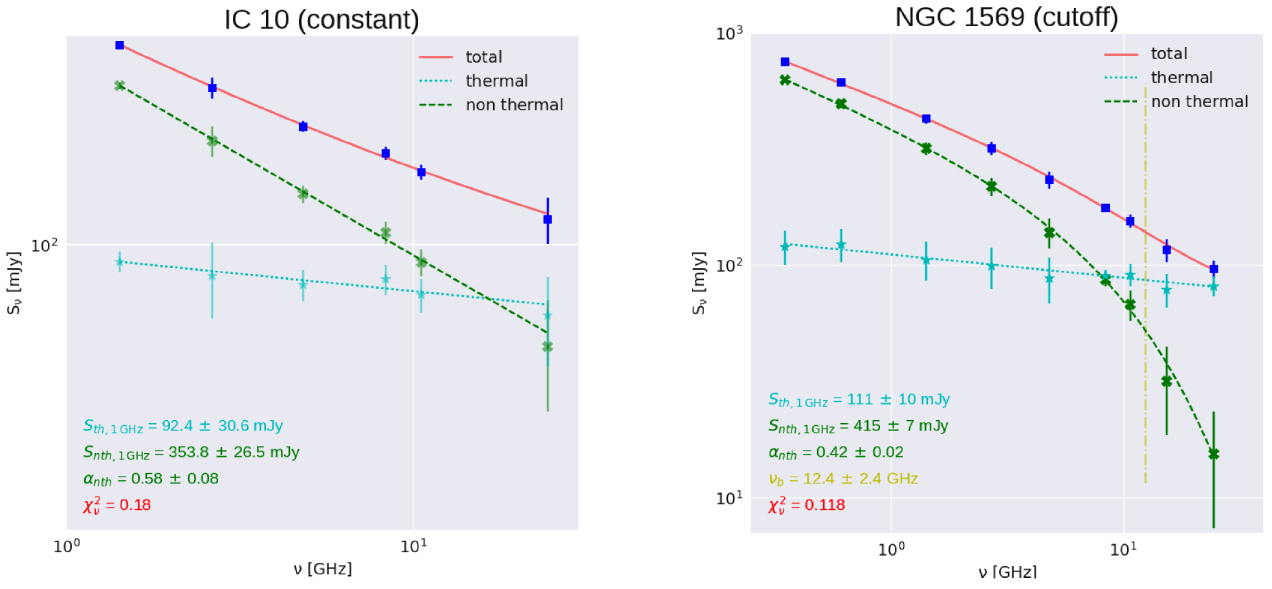}
\caption{Global radio continuum spectra from IC\,10 and NGC\,1569, obtained by the Effelsberg 100 m telescope \cite{85}. Measured flux densities:
blue squares connected by a red solid line; free-free emission: cyan stars connected by a cyan dotted straight line; non-thermal emission: green 
crosses connected by a green dashed line. Characteristic parameters are plotted in the lower left, indicating the thermal and non-thermal 
contributions at 1\,GHz. The vertical yellow dash-dotted line in the right panel denotes the break frequency (12.4\,GHz), which is also given 
in the lower left of this panel. The nonthermal spectral indices at 1\,GHz and the reduced $\chi^2$ values of the fits are presented in the lower 
left corners of both panels.}
\label{fig5}
\end{figure}

Ionized gas is observed in various parts of the electromagnetic spectrum and can comprise, in dwarf galaxies, a significant fraction of the entire 
ISM mass (e.g., \cite{84}). Starting with the lowest frequencies and longest wavelengths, ref. \cite{85} revealed, after disentangling global thermal 
and non-thermal radio emission between $\approx$345\,MHz and 24.6\,GHz, that the synchrotron emission of most dwarf galaxies (the exception is IC\,10) 
cannot be simulated with a single spectral index. While at lowest frequencies, an average spectral index of $\alpha$ = $-$0.59 $\pm$ 0.20 ($S_{\nu}$ 
$\propto$ $\nu^{\alpha}$; $S$: flux density) has been found, the synchrotron spectrum becomes steeper somewhere between 1 and 12\,GHz, due to a break 
or an exponential decline (\mbox{Figure~\ref{fig5}}). As a consequence, the free-free emission is stronger than previously anticipated, at least at GHz 
frequencies (see also Section \ref{S3.3}). The non-thermal cutoff could be caused by the injection spectrum of cosmic ray electrons, and would then 
be consistent with the mean spectral index of supernova remnants \cite{86}.

Within the local volume ($D$ $\la$ 11\,Mpc) already discussed in Section  \ref{S3.1} with respect to H{\sc i}, H$\alpha$ has been detected in 88\% of the 
379 late-type dwarfs, while the early-type dwarfs only show detections in 41\% (50/122) of the targets \cite{17}. Again this indicates that the 
late-type dwarfs tend to form stars more actively.

Finally, it is worth to mention the detection of [Ne\,V] in various dwarf galaxies, likely indicating fast shocks with speeds of several hundred
km\,s$^{-1}$, even though the presence of a massive compact nucleus is another possible interpretation \cite{87,88,89}. X-ray emission from BCDs is 
commonly dominated by binary stars, providing a component of intense hard radiation (e.g., \cite{90}).

\section{Star Formation and Active Nuclei} \label{S4}

\subsection{Star Formation} \label{S4.1}

A major question is how stars are formed in small galaxies. The Milky Way shows continuous star formation including many active sites at 
a given epoch. And while the locations of these star-forming regions may change with time, their number and SFR, integrated over the entire 
giant body of the galaxy, remained approximately constant during the last few Gyr (e.g., \cite{91,92}). Dwarfs appear to be different. The star 
formation history of Local Group dwarfs, shown by \cite{28}, involves spikes. While each galaxy has an individual star formation history, different 
from all the others, those galaxies with more recent star formation activity (which can be resolved in time much better than any such activity 
in the early days of the Universe) show brief periods of enhanced activity, clearly following the expectation of a bursty nature of star formation 
in small galaxies \cite{93,94,95}.

{\it GALEX}, the GALaxy Evolution eXplorer, was designed to observe the sky at ultraviolet (UV) wavelengths, 0.135 $\upmu$m--0.28 $\upmu$m.  According
to \cite{17}, 91\% (430/474) of the late type dwarfs and almost all BCDs inside of $D$ = 11\,Mpc were detected by {\it GALEX}, while the corresponding 
number for early-type dwarfs is much lower (28\%; 81/294). These values are consistent with the previously mentioned findings related to H{\sc i} 
and H$\alpha$ detection rates and emphasize the presence of a richer ISM related to massive star formation in the late-type sources. A comparison 
of far ultraviolet deduced star formation rates of late type dwarfs ($T$ = 9--10) and spirals ($T$ = 5--8) shows that present day star formation 
rates are compatible with the total stellar mass observed after 13.7\,Gyr. This holds for both late type dwarfs and spirals. With $M$(H{\sc i})/$M_*$ 
close to unity in late-type dwarfs (Section \ref{S3.1}), it is not surprising that a similar correlation also holds when comparing for this type of 
galaxies the FUV deduced star formation rate not with the stellar but with the H{\sc i} mass. The more crowded the environment, the larger is the 
scatter in sSFR = SFR/M$_*$, i.e., the specific star formation rate, because there are more galaxies that only show little star formation and 
are characterized by a low sSFR.

Star-formation rates from FUV luminosity refer to an equilibrium timescale of approximately 10$^8$\,yr (e.g., \cite{96}). SFRs from local dwarf 
galaxies ($\approx$900 targets; $D$ $<$ 11\,Mpc), based on FUV emission, exceed the average SFR over the age of the Universe by not more than an 
order of magnitude and reach this value only in exceptional cases \cite{17}. sSFR values are not exceeding sSFR = 10$^{-9.2}$\,yr$^{-1}$ versus an 
average of 10$^{-10.1}$\,yr$^{-1}$ over a Hubble time. {\it Hubble Space Telescope (HST)} data of asymptotic giant branch and helium burning stars 
analyzed by \cite{97} (23 targets) reveal highly individual star formation histories, like fingerprints, during the last 10$^8$\,yr. For about 
half of them also a spatial progression of star formation is found (see \cite{98} for a similar situation in dwarfs of the M\,81 group). Major bursts 
beyond a factor of three over the average during the studied time interval ($\approx$10$^8$\,yr) are not encountered. This is in line with \cite{96}, who 
analyzed the H$\alpha$ (characterizing a timescale of $\approx$5 $\times$ 10$^6$\,yr) versus ultraviolet emission (timescale: $\approx$10$^8$\,yr) 
of 185 local galaxies. Dwarfs near the upper end of our luminosity limit show burst amplitudes (with respect to their average SFR) not exceeding 
a factor of ten, while galaxies below $M$ = 10$^{7.5}$\,M$_\odot$ may undergo large and fast bursts, reaching an amplitude of up to two 
orders of magnitude over the average SFR, within a timescale of $t$ $<$ 30\,Myr. 

H$\beta$ and UV luminosities should be considerably enhanced during the short periods of ongoing starbursts, with the H$\beta$ emission decreasing
more rapidly than the UV emission after the peak of the burst. This is because H$\beta$ is emitted by younger and more massive stars. Therefore 
determining the UV and H$\beta$ luminosities, both values can be extrapolated to a zero age, where the starburst was at its peak. Using these 
extrapolated peak values instead of the directly measured luminosities based on $\approx$14,000 compact star-forming galaxies at redshifts 
$z$ $<$ 1, ref. \cite{4} succeeded in significantly reducing the scatter in the sSFR values, which is found to be independent of galactic stellar 
mass, further supporting the notion of a bursty mode of star formation in galaxies of small size, mass and luminosity. 

During such a burst of star formation, observations suggest the presence of two modes: a more `passive' and a more `active' one. In the latter
mode, the BCDs form SSCs, show high column densities of H$_2$ and dust, including high dust temperatures. All this is confined to compact ($<$100\,pc) 
regions. The less extreme mode shows instead more widespread and more diffuse gas with lower column densities \cite{74,99}. Characteristic examples 
showing these two modes (e.g., SBS\,0335--052 versus IZw\,18; \cite{84}) are discussed in some detail in \mbox{Section \ref{S5.2}.}

\subsection{Combining Different Interstellar Tracers} \label{S4.2}

While we have already combined H$\alpha$ or H$\beta$ and UV emission to tackle the bursty nature of star formation in galaxies of low mass (see the last 
section), it is also noteworthy that at metallicities 12 + log(O/H) $<$ 8.2, not only CO but also PAH emission becomes more and more elusive (e.g.,
\cite{74,100}; Section  \ref{S3.2}). This is only indirectly related to metallicity and primarily due to a hard, intense radiation field \cite{100}. 
Based on {\it ALMA} measurements of the Local Group dwarf NGC\,6822 with a 2\,pc linear resolution, ref. \cite{77} finds that on pc-scales CO 
emission correlates well with PAH emission and less well with 24 $\upmu$m emission. An anticorrelation is found with H$\alpha$. Thus PAH emission 
may be a possible alternative to CO as a tracer of molecular gas.

Using stacked averages of 1.4\,GHz radio continuum and IR-70 $\upmu$m Spitzer data, ref. \cite{98} finds that the correlation between these two SFR tracers, 
observed in large galaxies on scales encompassing at least several 100\,pc, also holds in dwarf galaxies. This may be the result of a `conspiracy': 
the underabundance of dust due to low metallicities may be compensated by an easier escape of the cosmic ray electrons from supernovae, 
that give rise to the 1.4\,GHz emission.

Characteristic H{\sc i} distributions in actively star-forming dwarf galaxies show large (100--1500\,pc) H{\sc i} holes due to stellar feedback. With 
H{\sc i} layers much thicker than in spiral galaxies (Section \ref{S2.2}) gas densities are comparatively low, allowing for the formation of large shock-driven 
shells. While the smaller ones are filled with H$\alpha$ emitting gas, the larger ones tend to show H$\alpha$ along their rims (e.g., \cite{101,102,103,104}). Typical 
expansion velocities are 5--10\,km\,s$^{-1}$, not fast enough to release the gas into the intergalactic medium unless the galaxy has a mass well below 
10$^7$\,M$_{\odot}$ \cite{105}. H$\alpha$ and thermal X-ray emission may be correlated on global scales but not locally (e.g., \cite{106}), with the hot gas possibly 
leaking out of the H{\sc ii} regions \cite{107,108}. 

In regions of low metallicity, CO is able to trace the densest and most shielded cores inside of gas agglomerations mainly consisting of H{\sc i} and 
H$_2$. However, it commonly only reveals a small fraction of the molecular environment. H$_2$ itself is also not providing a clear picture because it is 
homonuclear, lacking a permanent dipole moment. Due to its light weight, relevant frequencies are shifted away from radio wavelengths into the infrared 
and excited states are only found at levels $>$100\,K above the ground state, too high to trace the bulk of the usually cooler molecular gas. Singly 
ionized carbon, i.e., the [C{\sc ii}] fine structure line at 158 $\upmu$m, can help to complement CO and H$_2$ observations, even though it is an atomic 
constituent. [C{\sc ii}]$\lambda$158 $\upmu$m is usually the strongest far-infrared line and thus, with respect to its intensity, a prominent coolant of 
the ISM. With the carbon ionization potential of 11.26\,eV, well below that of hydrogen, 13.60\,eV, singly ionized carbon is a multiphase tracer, 
including diffuse gas in the warm and cold neutral ISM, the warm ionized ISM and dense gas in photon dominated regions. While $L$([C{\sc ii}])/$L$(CO 
$J$ = 1$\rightarrow$0) $\approx$ 2000 in normal spiral galaxies and about twice as large in starburst environments of major galaxies, ratios tend to be 
much higher, up to 80,000, in dwarfs of low metallicity (e.g., \cite{55}). This is accompanied by a milder increase in the [C{\sc i}] to CO 1$\rightarrow$0 
ratio \cite{109}.

With [C{\sc ii}] potentially arising from ionized, neutral atomic and neutral molecular gas, the main question is how much these physically different 
environments can contribute to the observed strong line emission. This ambiguity constitutes the challenge of using this tracer to probe CO dark 
molecular gas. To give an example: Studying five positions in NGC\,4214 (12 + log(O/H) $\approx$ 8.2) belonging to three separate regions with 
beam sizes of 12--14$^{\prime\prime}$ and using ancillary data from far infrared fine structure lines, H{\sc i} and CO $J$ = 1$\rightarrow$0 and 
2$\rightarrow$1, ref. \cite{110} finds that about 80\% of the total molecular hydrogen mass is CO-dark, i.e., not participating in CO emission. 
The method to quantify the contribution of the CO-dark molecular gas with respect to the entire molecular environment (e.g., \cite{48})  may include 
a detailed comparison between the lineshapes of [C{\sc ii}]\,158 $\upmu$m, [N{\sc ii}]\,122 $\upmu$m and [N{\sc ii}]\,205 $\upmu$m, observable with 
{\it Herschel}, and H{\sc i} and CO to kinematically discriminate between the various gas components. The observed features have then to be used to 
obtain estimates of column densities as a function of radial velocity. Assumptions on fractional abundances relative to hydrogen complete this process. 
In particular the latter steps related to the evaluation of column densities and fractional abundances lead to substantial uncertainties. Therefore, 
high quality data providing information on crucial parameters as a function of velocity are mandatory.  

To summarize, $L$([C{\sc ii}])/$L$(CO $J$ = 1$\rightarrow$0) luminosity ratios are largest in dwarf galaxies because the CO-dark gas is even more widespread 
there than in larger more metal-rich galaxies. The [C{\sc ii}] 158 $\upmu$m line has become the main tracer of this CO-dark gas, although there may be exceptions 
(for one such exception, IZw\,18, see \cite{111}). The CO-dark gas apparently dominates the molecular environment of dwarf galaxies and [C{\sc ii}] can even be 
used to estimate $X$(CO) and $\alpha$(CO) values (\cite{112}; see also Section \ref{S3.4}) and to test the Schmidt-Kennicutt star-formation law \cite{113}. The 
latter relation connecting molecular and star-forming surface densities appears to be consistent with that of large galaxies \cite{113,114}, although the metallicity 
certainly plays a role in the fractional amount of the CO-dark mass. Nevertheless, the primary factor regulating this quantity is almost certainly the extinction 
characterizing the dense gas \cite{48}.

Finally, much effort has also been spent on chemical abundances, and the combination of tracers of different elements. Optical and near-infrared spectroscopy 
of H{\sc ii} regions allows for the detection of a plethora of lines from a number of abundant heavy elements that are not only constraining stellar evolution 
and the `chemical' evolution of dwarf galaxies, but also the depletion of metals onto dust grains. While the latter effect is not as severe as in more metal-rich 
galaxies with higher dust-to-gas mass ratios (see also Section \ref{S2.2}), the multitude of elements to be detected leads to a large number of notable correlations. To 
mention a few: [Mg/Ne] decreases with rising metallicity because Ne, unlike Mg, is a noble gas which is not significantly incorporated into dust grains. On the 
other hand, Mg and O depletion is less metallicity dependent than Fe depletion and data including SBS\,0335--052, a galaxy with a particularly low metallicity 
close to 12 + log(O/H) = 7.0, indicate that C/O increases with O/H \cite{32}. However, in view of the many elements involved, a detailed description of these relations 
is beyond the scope of this article. For details, see e.g., \cite{32,58,115,116}.

\subsection{Active Galactic Nuclei and Outflows} \label{S4.3}

Active galactic nuclei (AGN) have the potential to shape their parent galaxy and to severely affect its ISM. For a long time, searches for AGN in small 
galaxies of low mass were unsuccessful. Nevertheless, this is an important endeavour, because the study of central black hole candidates in dwarf galaxies 
would provide strong constraints to supermassive black hole (SMBH) seeds in the early Universe. In recent years, evidence for the presence of massive black 
holes is gradually growing. The low end of the black hole mass - velocity dispersion relation in Seyfert~1 and Seyfert~2 galaxies has been studied by 
\cite{117,118,119}. The detection of AGN in five dwarf galaxies based on the presence of extraordinarily broad and luminous H$\alpha$ lines has been 
suggested by \cite{120,121}. The observed emission, with line widths in excess of 2000\,km\,s$^{-1}$ (see also \cite{122}), could not be explained by 
supernovae, Wolf-Rayet stars or shocks propagating in circumstellar envelopes, either because these objects would not provide the observed high luminosities 
or because of a lack of variability over a time span of order 5\,yr. An AGN is in this context the remaining, most likely explanation. 

Employing adaptive optics from the {\it Gemini}/NIFS instrument complemented by {\it HST} imaging,  ref. \cite{123} reported the likely discovery of supermassive 
nuclear engines ($M$ $>$ 10$^6$\,M$_{\odot}$) in two ultracompact dwarf (UCD) galaxies of the Virgo cluster. This is consistent with the idea that such
objects represent stripped nuclear regions of previously more extended galaxies. Based on the discovery of a hard X-ray source at the dynamical center 
of Henize\,2--10, \cite{124,125} reported the presence of an accreting black hole in this galaxy, surrounded by forming super star clusters. Following \cite{126}, 
the mass of the nuclear object is $\approx$ 3 $\times$ 10$^6$\,M$_{\odot}$, the uncertainty being a factor of a few. Nevertheless, the absolute magnitude 
of this galaxy is with $M_{\rm V}$ $\approx$ $-$19$^{\rm M}$ slightly above the luminosity limit we defined for dwarfs in Section \ref{S1}. Another candidate 
galaxy, with similar visual luminosity, is NGC\,5408, where \cite{127} suggested the presence of a nuclear source based on X-ray, optical and radio observations. 
While the X-ray emission from these sources could serve as a model, ref. \cite{128} did not detect strong hard UV or X-ray emission from the putative accretion 
disks around candidate black holes in small galaxies that show particularly strong and wide emission lines of ionized gas. 

Using the Baldwin-Phillips-Terlevich BPT diagram \cite{129} and connecting [O{\sc iii}]/H$\beta$ with [N{\sc ii}]/H$\alpha$ line intensity ratios, ref.
\cite{130} identified among $\approx$2500 SDSS targets 136 dwarf galaxies with line ratios indicating the presence of an AGN. A small fraction of 
these sources also shows broad H$\alpha$ emission. Using the virial theorem, the median mass of the putative AGN is estimated to be $M$ = 
2 $\times$ 10$^5$\,M$_{\odot}$. It should again be noted, however, that the median absolute magnitude of these objects is $M_{\rm g}$ $\approx$ 
$-$18$^{\rm M}$, so that a significant fraction of these galaxies is more massive than the true dwarfs discussed in this article (\mbox{Section \ref{S1}}). 
Addressing galaxies with lower masses more consistent with our definition of dwarfs, ref. \cite{131} detected [Fe\,X]$\lambda$6374 coronal line emission in 81 
galaxies with $M_*$ $<$ 3 $\times$ 10$^9$\,M$_{\odot}$. This emission is found to be too strong to be emitted by stellar sources and even supernovae 
are not able to explain the majority of the observed lines. 

To summarize, the systematic study of massive black holes in dwarf galaxies has just begun. It will be interesting to see the impact of these studies on our 
understanding of formation of and feedback from massive nuclear sources in the early days of the Universe \cite{131,132,133,134,135,136,137,138,139,140}.

\section{Star Forming Dwarfs in the Local Universe} \label{S5}

\subsection{Are There Young Dwarfs in the Local Universe?} \label{S5.1}

Analyzing $\approx$280,000 and 400,000 SDSS spectra, ref. \cite{141,142} searched for galaxies that may have formed most ($\ga$50\%) of their stellar 
component during the last Gyr, while lacking an active galactic nucleus. Because of a lower mass cutoff at $M$ = 10$^8$\,M$_{\odot}$, 
metallicities (12 + log(O/H) = 7.9 -- 8.6) are at the upper end of what is being considered here (Section \ref{S1}). When compared with a control 
sample involving the same selection criteria but requiring a predominantly older stellar population, the potentially young galaxies are bluer, 
have higher sSFRs and surface brightnesses, and are relatively dusty, compact, asymmetric and clumpy. Furthermore, they are with respect to 
H{\sc i} significantly more gas rich and tend to reside in the inner parts of low-mass groups. Appearing to be relatively frequently affected by 
interactions, they may have been activated by such an event. Modelling galaxy formation and trying to simulate the number of such young galaxies 
in the local Universe, ref. \cite{143} provides predictions for a wide range of scenarios. These, including or excluding dark matter, yield 
results that differ by several orders of magnitude, emphasizing that the systematic observation of young local galaxies could become an important 
constraint to galaxy formation and the evolution of the Universe.

There are galaxies with particularly low gaseous metallicities, reaching 12 + log(O/H) = 7.0 in the local volume ($z$ < 1.0) \cite{144,145,146,147}. 
These galaxies may be affected by particularly inefficient star formation and/or severe losses of metal enriched matter due to supernovae or large 
scale galactic winds. These galaxies also contain a predominantly young stellar population. Investigating large samples of galaxies from the SDSS, 
the minimal metallicities encountered in these galaxies provide, like the above mentioned number of potentially young galaxies, important 
clues for the formation and evolution of galaxies during the last 13 billion years.

Finally, it is worth mentioning a substantial lack of almost purely gaseous agglomerations on galaxy scales in the local Universe. An area of 
36\,deg$^2$ was surveyed in H{\sc i} with the 64\,m {\it Parkes} multi-beam system \cite{6}, with 129 sources detected. The same area was also 
observed with the {\it UK Schmidt Telescope} reaching a limiting surface brightness of $\mu_{\rm R}$ = 26\ffm5\,arcsec$^{-2}$; no optically dark 
H{\sc i} galaxies were found. Such galaxies must either be rare, gas poor or otherwise outside their detection limits.

\subsection{Some Outstanding Targets} \label{S5.2}

In the following we discuss properties of local individual sources that illustrate the large variety of metal-poor actively star forming 
dwarf galaxies, on the basis of what has been described in previous sections. This will be complemented, in \mbox{Section \ref{S6}}, by dwarf
galaxies at cosmologically relevant distances. The chosen order of the sources in the following subsection does not reflect physical properties 
but merely the right ascension of the targets. 

{\bf IC\,10 (UGC\,192)}: IC\,10 is commonly classified as a Local Group dwarf iregular at a distance of $\approx$800\,kpc with $M_{\rm B}$ = 
$-$16$^{\rm M}$\!\!\!.\,7 and an oxygen abundance about 1/4 solar (\mbox{e.g., \cite{148}}). It is, like NGC\,4449, one of those few galaxies with 
an H{\sc i} envelope that extends far beyond its optical diameter, encompassing an angular size of a full degree \cite{149,150,151,152,153}. The 
velocity gradient along the major axis is opposite to that in the core of the galaxy, possibly indicating a non-planar highly warped H{\sc i} disk, 
again similar to that seen in NGC\,4449. There are holes in the H{\sc i} distribution, likely a consequence of supernovae or stellar winds \cite{154}. 
Modeling the ionized gas through observations of mid- and far-infrared fine structure lines and the photoionizaton code CLOUDY, ref. \cite{155} finds 
that most ionized clumps, irrespective of their size, show hydrogen densities of order 10$^{2.0}$ to 10$^{2.6}$\,cm$^{-3}$ and ages of about 5\,Myr. 
The larger sized clumps are more radiation bounded than the smaller ones, while [C{\sc ii}], [Si{\sc ii}] and [Fe{\sc ii}] emission mostly 
originates from the neutral gas.

Most of the star-forming activity of this galaxy inside the Local Group is taking place in an elongated 3$^{\prime}$ $\times$ 1\ffam5 sized region 
extending from the northwest to the southeast, as seen in 2MASS 2 $\upmu$m emission \cite{148} and in the $\approx$10\,GHz radio continuum \cite{156}. Two 22\,GHz 
H$_2$O masers are found in this region \cite{157,158}, indicating ongoing vigorous star formation. This is complemented by the presence of massive 
molecular clouds traced by CO \cite{157,159,160}. The emission includes 14 Giant Molecular Clouds (GMCs) with sizes, CO luminosities, line widths and 
X(CO) conversion factors similar to the values found in the Galactic disk. The star-forming rate per molecular mass is higher than in most other 
galaxies, confirming that IC\,10 is presently undergoing a starburst. 

{\bf NGC\,147 (UGC\,326), NGC\,185 (UGC\,396) and NGC\,205 (UGC\,426)}: The three dwarf ellipticals with absolute visual magnitudes of $M_{\rm V}$ = 
$-$15$^{\rm M}$\!\!\!.\,0, $-$15$^{\rm M}$\!\!\!.\,0 and $-$16$^{\rm M}$\!\!\!.\,2 (NED\footnote{NASA/IPAC extragalactic database, ned.ipac.caltech.edu}), 
respectively, are companions of the Andromeda galaxy and are therefore nearby, offering the possibility to study their ISM in spatial detail. 
In line with the highly individual star formation histories of Local Group dwarfs already mentioned in Section \ref{S4.1}, they offer a useful insight into 
the variety of properties one may encounter when analyzing galaxies that appear to be similar at first sight. This holds in particular for NGC\,147 
and NGC\,185: Luminosity, mass, metallicity, and size are extremely similar. However, NGC\,147 does not possess a significant ISM, while NGC\,185 
does, and also contains a young stellar component (e.g., \cite{32}). NGC\,205 does not only host an ISM and a young stellar component, it is also 
nucleated, with a stellar cluster at its core {(e.g., \cite{161})}. Its neutral atomic and molecular ISM, comprising within a factor of a few $M$ $\approx$ 
10$^7$\,M$_{\odot}$ \cite{145}, shows similar morphologies and kinematics on 90\,pc scales \cite{162,163}. There are attempts to reveal a massive nuclear engine 
at its center (e.g.,  \cite{164,165}). So far, based on stellar dynamics, a value of $M_{\rm BH}$ $\sim$ 7 $\times$ 10$^3$\,M$_{\odot}$ has been proposed  
by these authors, both as a 3$\sigma$ upper limit and as an actual value. However, in view of systematic errors and the possibility of a cluster of dark 
stellar remnants, this can only be considered as tentative.

\begin{figure}[H]	
\includegraphics[width=13.5cm]{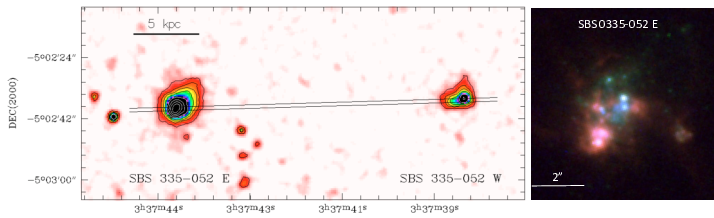}
\caption{\textbf{Left}: A B-band image taken from the {\it Calar Alto} observatory showing the western and eastern components of SBS\,0335--052. The black lines 
indicate the location of a slit used for optical and near infrared spectroscopy. Taken from \cite{60}. \textbf{Right}: A composite true-color HST/ACS image of 
SBS\,0335--052 that  illustrates the super star clusters in the eastern source. Taken from \cite{166}.}
\label{fig6}
\end{figure}

\begin{figure}[H]	
\includegraphics[width=13.5cm]{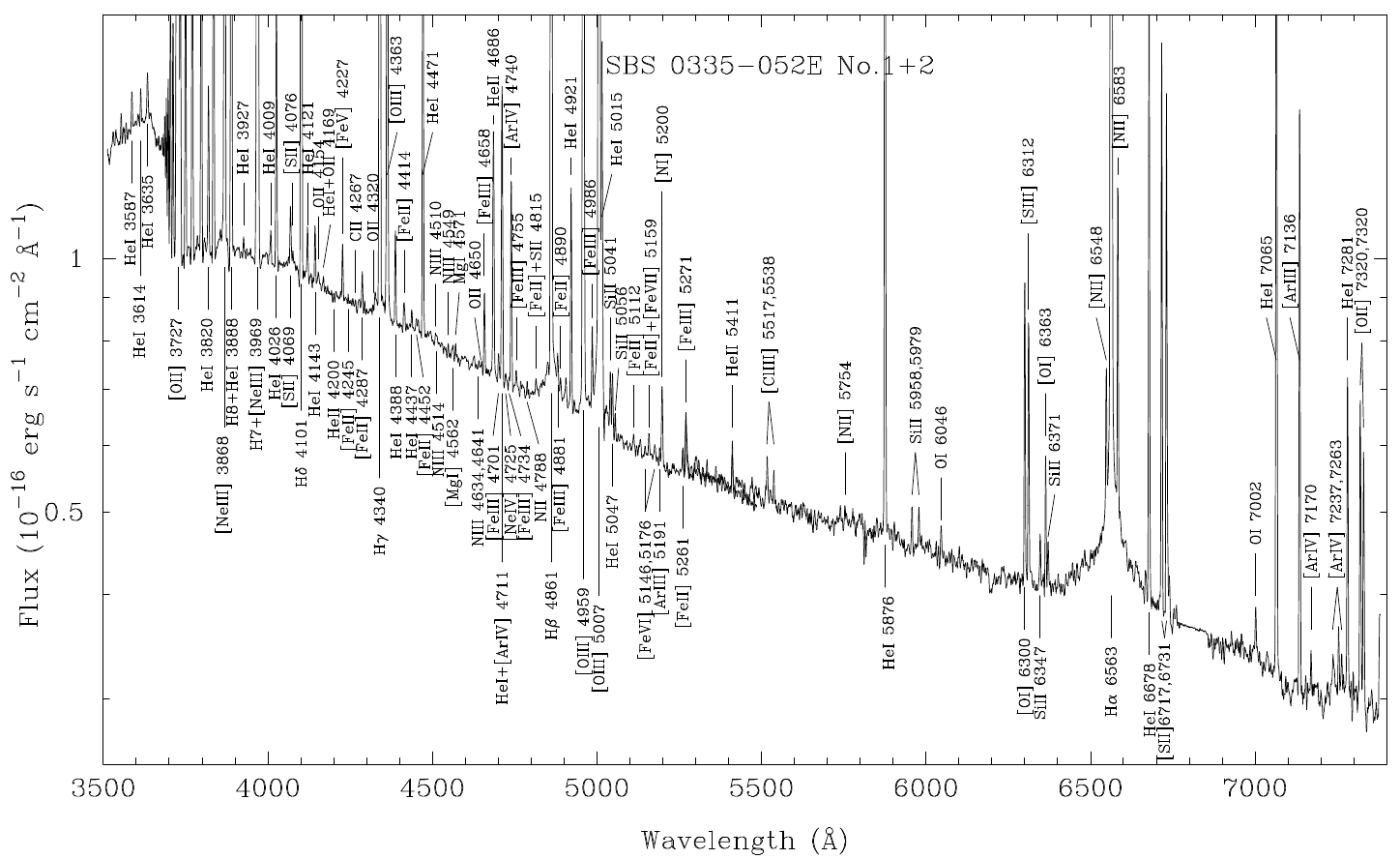}
\caption{A FORS high-resolution spectrum taken with the {\it Very Large Telescope (VLT)} of the European Southern Observatory (ESO) from the two southern 
SSCs of SBS\,0335-0052 (see Figure~\ref{fig6}, right panel), employed to derive metallicities. The spectrum, showing particularly wide hydrogen recombination lines,
was published by \cite{60}.}
\label{fig7}
\end{figure}

\begin{figure}[H]	
\includegraphics[width=6.5cm]{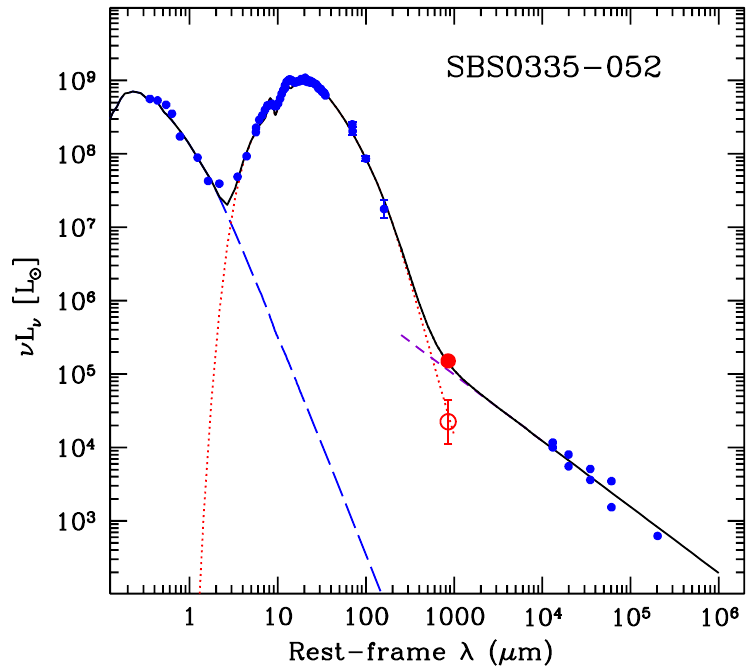}\includegraphics[width=6.5cm]{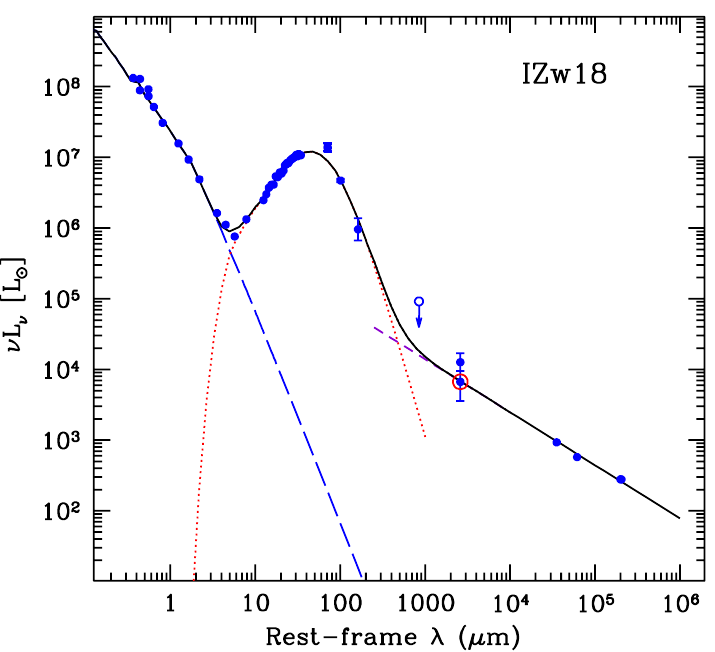}
\caption{Spectral Energy Distributions (SEDs) from SBS\,0335--052 (\textbf{left}) and IZw\,18 (\textbf{right}). The black lines respresent fits using DUSTY models, the 
blue dashed lines show the stellar contributions, the dotted red lines give the contributions from the dust and the dashed purple straight line (mostly 
undistinguishable from the total emission) provides the radio continuum at long wavelengths. \textbf{Left}: The filled red circle indicates the total $\lambda$ 
= 850 $\upmu$m continuum emission from 0\ffas7 $\times$ 0\ffas45 {\it ALMA} data, while the empty red circle stands for the dust contribution. \textbf{Right}: The 
empty red circle surrounds the dust contribution suggested by \cite{171}, while the blue filled circle above includes the entire $\lambda$ = 2.6\,mm emission. 
From \cite{51}.}
\label{fig8}
\end{figure}

{\bf SBS\,0335--052 and IZw\,18 (Mrk\,116)}: The first target consists of a pair of two BCDs, SBS\,0335--052W and SBS\,0335--052E (see Figures~\ref{fig6} and \ref{fig7}), 
embedded in a common H{\sc i} \mbox{cloud \cite{167,168}} and exhibiting vigorous star formation. This includes the presence of young SSCs, possibly a consequence 
of their interaction. The most interesting aspect of this system ($D$ $\approx$ 55\,Mpc; $M_{\rm V}$ $\approx$ $-$16$^{\rm M}$\!\!\!.\,8, \cite{169}) is its 
extremely low metallicity. For the brighter eastern source \cite{60} find 12 + log(O/H) = 7.11 -- 7.32 in different H{\sc ii} regions. In the fainter, western 
one projected 22\,kpc away from SBS\,0335--052E they obtain for the main source 12 + log(O/H) = 7.22 $\pm$ 0.07. However, in three less conspicuous ionized
bubbles, the metallicity is exceedingly low, with  values of 7.01 $\pm$ 0.07, 6.98 $\pm$ 0.06 and 6.86 $\pm$ 0.14. Thus SBS\,0335--052 is a clear candidate for 
the class of 'young dwarfs' discussed in Section \ref{S5.1}. The spectral energy distribution (SED) of SBS\,0335--052E (see Figure~\ref{fig8}) covering six orders of 
magnitude in wavelength shows the typical peaks due to stellar emission near optical wavelengths and the dust emission in the infrared, as well as additional 
free-free emission at longer wavelengths \cite{51}. Noteworthy is that even at 870 $\upmu$m ($\approx$345\,GHz) almost 90\% of the emission appears to represent 
free-free emission, likely because the ISM is affected by (1) a high level of star formation and (2) a low dust-to-gas mass ratio (see also Section \ref{S3.3}). 
CO remains undetected \cite{170}.

\vspace{-12pt}

Following the definition in Sect.\,5.1, IZw\,18 might be another young galaxy, although this is  matter of debate because of the presence of evolved stars 
\cite{172,173}. It also hosts H{\sc ii} regions of extremely low metallicity, 12 + log(O/H) = 7.17 $\pm$ 0.03 in the northwestern and 7.18 $\pm$ 0.03 in the 
southeastern component \cite{174}. Showing extremely weak CO \mbox{emission \cite{175}}, the galaxy may still be in the process of formation \cite{176}. IZw\,18 is with 
$D$ $\approx$ 18\,Mpc ($M_{\rm V}$ $\approx$ $-$15$^{\rm M}$\!\!\!.\,0; NED) much closer to us than SBS\,0335--052. No red giant is found in measurements 
1--2 magnitudes deeper than the expected tip of the red giant branch. Many stars appear to be younger than half a billion years, a time during which 
up to three bursts of star -formation can be deduced, interrupted by more quiescent periods lasting 1--2 $\times$ 10$^{8}$\,yr.  Far UV (917--1188\,\AA) 
absorption lines \cite{177} do not indicate the existence of dominant galaxy-wide in- or outflows. Based on {\it HST} far UV absorption line data,  \mbox{ref. \cite{50}} 
investigated the metallicity of the massive H{\sc i} envelope of IZw\,18. It is as in SBS\,0355--052~\cite{168} not zero, but roughly two thirds of that derived 
from the H{\sc ii} regions. They argue that at least a part of this metallicity must have been generated prior to the present starburst, suggesting the 
existence of an older stellar generation.

Comparing SBS\,0335--052 with IZw\,18, these two systems with extremely low metallicities may be good examples of the two starburst modes in dwarfs suggested
earlier at the end of Section \ref{S4.1}. SBS\,0335--052 would then represent the `active' and IZw\,18 the `passive' mode. The infrared luminosity of SBS\,0335--052 is 
with $\approx$1.6 $\times$ 10$^9$\,L$_{\odot}$ versus $\approx$2 $\times$ 10$^7$\,L$_{\odot}$ for IZw\,18 almost two orders of magnitude higher, indicating quite 
different levels of star-formation activity \cite{51}. At {radio wavelengths, the two galaxies show very different spectra: while IZw\,18 has a relatively 
straight power-law indicative of optically thin thermal emission combined with synchrotron radiation, SBS\,0335--052 is self-absorbed (e.g., \cite{178,179,180}). This 
indicates ionized gas densities of 3000 to 5000\,cm$^{-3}$, much higher than the implied densities of $<$100\,cm$^{-3}$ in IZw\,18 \cite{181}. Another major difference 
between the two galaxies is the peak position of their IR-dust SEDs. Both show maxima at short wavelengths, compatible with the relatively high dust temperatures 
characterizing dwarf galaxies (Section \ref{S3.2}). However, while IZw\,18 peaks at 50--60 $\upmu$m, SBS\,0335--052 is truly extreme showing its maximum in the range 
20--30 $\upmu$m (Figure~\ref{fig8}), thus indicating a lack of a cool dust component. In view of these differences we conclude that metallicity is not the only factor 
that shapes dwarf galaxies with an extremely unprocessed interstellar medium.

\begin{figure}[H]	
\includegraphics[width=11.5cm]{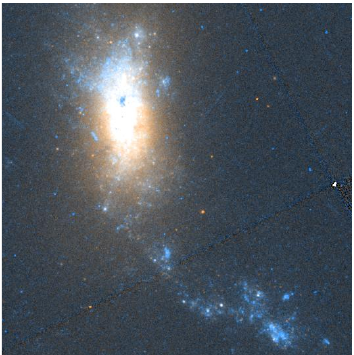}
\caption{A composite {\it HST} image of NGC\,1140, taken from \cite{74}.} 
\label{fig9}
\end{figure}

{\bf NGC\,1140 (Mrk\,1063)}: NGC\,1140 (Figure~\ref{fig9}) is a barred irregular galaxy at a distance of $\approx$20\,Mpc with a metallicity of about 12 + 
log(O/H) = 8.2 \cite{74}. Its irregular morphology, revealed by H$\alpha$ images \cite{182} shows a southwestern chain of H{\sc ii} regions, the tail, and a 
northern region, the head, which dominates the emission and star forming activity. This head, with 30 times the H$\alpha$ luminosity of the giant 
H{\sc ii} region 30\,Dor in the LMC \cite{183}, is surrounded by less conspicuous star forming regions. The dominant complex, about 500\,pc in size, 
contains 6--7 blue SSCs \cite{183,184}, the most luminous of them consisting of several thousand O stars \cite{185}. There is no galactic scale outflow, but 
violent shocks from the SSCs are disrupting the ISM on scales of 1--2\,pc \cite{182}. Shape and activity of the galaxy may be caused by a merger 
\cite{182,183}. 

As in NGC\,5253 (see below), CO is highly excited and not optically thick, with $J$ = 3$\rightarrow$2 line flux densities far surpassing those of the 
$J$ = 1$\rightarrow$ 0 line. Naively, one would therefore expect high kinetic temperatures of the molecular gas giving rise to the observed CO emission. 
However, model fits including the dust continuum and molecular and atomic abundances relative to hydrogen \cite{109}  reveal high densities ($n$(H$_2$) 
$>$ 10$^5$\,cm$^{-3}$) but low temperatures ($T_{\rm kin}$ $\approx$ 20\,K), CO/C{\sc i} $\approx$ 0.1, $^{12}$CO/$^{13}$CO $\approx$ 8--20 and 
$N$(CO) $\approx$ 5 $\times$ 10$^{16}$\,cm$^{-2}$ for a realistic gas-to-dust mass ratio of $\approx$380 \cite{109}. The low kinetic temperature in combination 
with a rather low opacity ($\tau$($J$ = 3$\rightarrow$2) $\approx$ 0.2) can be explained if we assume that CO is only existing in spatially highly confined 
particularly well shielded pockets, where irradiation by UV photons is not as strong as elsewhere in the complex. Overall, the CO fractional abundance 
turns out to be $\approx$8 $\times$ 10$^{-7}$, two orders of magnitude below the commonly encountered value (HCN remains undetected). This is similar to 
the situation in the massive star forming region N159W of the LMC, where ammonia (NH$_3$), overall also highly underabundant, provides a temperature of 
only $\approx$16\,K \cite{186}, while H$_2$CO, a molecule which is less sensitive to ionizing radiation, shows kinetic temperatures of order 50\,K \cite{187,188}. 
Presumably, there is no chance to detect NH$_3$ in NGC\,1140. Here CO plays the role NH$_3$ is playing in the LMC because of its
sensitivity to the strong ionizing radiation fields at low metallicity. 

The extremely low $^{12}$CO/$^{13}$CO abundance ratio obtained by \cite{109} is smaller than any value encountered in the ISM of the Milky Way (see \cite{189} 
for an analysis of the region with the lowest Galactic values, the Central Molecular Zone (CMZ) of the Galaxy). It does not reflect the real $^{12}$C/$^{13}$C 
isotope ratio but must be caused by fractionation, i.e., by chemical charge exchange reactions that can reduce the $^{12}$CO/$^{13}$CO abundance ratios to 
values below 20 (Figure~\ref{fig10}), even if the actual $^{12}$C/$^{13}$C isotope ratio is similar to that in the LMC, i.e., of the order of 55.

\vspace{-12pt}
\begin{figure}[H]	
\includegraphics[width=11.5cm]{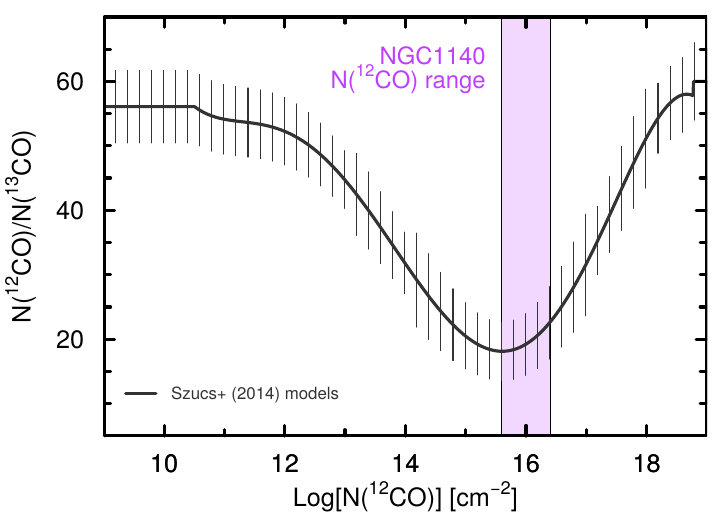}
\caption{The carbon isotope ratio $^{12}$C/$^{13}$C as traced by CO is plotted against the CO column density. The degree of fractionation follows
the models of \cite{190}; taken from \cite{109}.}
\label{fig10}
\end{figure}

{\bf NGC\,5253}: At a distance of about 4\,Mpc and with a metallicity of the order of 12 + log(O/H) = 8.25 \cite{191,192}, the BCD NGC\,5253 is one of 
the best examples to study the birth and evolution of SSCs as well as the interaction between densely packed massive stars and the surrounding gas. 
Not unexpected in view of the SMC-like metallicity of NGC\,5253 (see Section \ref{S3.4}) the {\it ALMA} data from \cite{192}, obtained with an angular resolution 
of 0\ffas2 ($\approx$3\,pc), reveal 118 individual CO clumps. The most prominent object associated with a nascent SSC is a cloud termed `D1', which 
contains within a diameter of only about 6\,pc around 1500--2000 O stars \cite{193}. Therefore it represents a promising candidate for the existence 
of a top heavy initial mass function (IMF). The correlation between CO line width and cloud size provides enhanced line widths by 0.2--0.3\,dex 
with respect to Galactic disk clouds in the most active regions. The ionized gas comprises (as in Mrk\,71, see Section \ref{S3.5} two components that are 
offset in velocity and in this case also in position \cite{194}. These {\it JVLA} observations at an angular resolution of 0\ffas15 also show an arm of ionized 
gas reaching out of the center in northwestern direction, possibly indicating feedback.

To further focus on D1, the most extreme source in the galaxy: the age of its SSC is estimated to be only $\approx$10$^6$\,yr, far too young for any 
supernova event \cite{195}. The numerous massive stars must create isolated bubbles of ionized gas. However, surprisingly, the effect of the ionizing 
radiation and the stellar winds of the numerous massive stars leave the parent CO cloud mainly undisturbed. Mass loading by a large number of 
associated lower mass stars downstream of the flows could enhance the cooling and confine the hot shocked winds from the massive stars \cite{196}.
Nevertheless, these hot shocked winds have to merge in the future, then providing a large scale galactic wind.

\section{Star Forming Dwarfs in the Cosmological Persepctive} \label{S6}

\subsection{Primordial Abundances}

Dwarf galaxies contain matter that is relatively unaffected by contamination from stellar ejecta. Therefore they are suitable targets to trace 
the abundances of those nuclei that are, according to the standard model, products of Big Bang nucleosynthesis. These nuclei are deuterium
(D), the stable rare isotope of helium ($^3$He), the main helium species ($^4$He), and the most common lithium isotope ($^7$Li). Abundances can 
be studied taking spectra of nearby objects with low metallicity. The two keystones to constrain Big Bang nucleosynthesis are deuterium and the 
main helium isotope. The former is most conveniently determined in damped Ly-$\alpha$ absorption systems towards quasars at high redshift, which 
is outside the scope of this article, while $^4$He is to date most accurately traced by sensitive observations of H{\sc ii} regions in local metal 
poor dwarf galaxies (e.g., \cite{197,198,199,200,201,202,203,204}). In such environments it is acceptable to assume that there is a linear correlation between helium and oxygen 
abundances, so that an extrapolation to an oxygen abundance of zero, the primordial value, can be carried out. Nitrogen or sulfur abundances may 
also provide a useful anchor. With respect to oxygen, it is important to include the weak [O{\sc iii}]$\lambda$4363\AA\ line, which requires regions 
of high electron temperature, potentially leading to a notable effect due to collisional excitation. To unravel the physical conditions of the gas,
the inclusion of the He\,I line at $\lambda$ = 10830\AA\ \cite{197} adds essential information, thus reducing errors in the helium abundance considerably. 
Combining the results of the above mentioned eight articles results in a helium abundance by mass of $Y_{\rm p}$ = 0.2462 $\pm$ 0.0013 relative to 
hydrogen. Weighting the values with the inverse square of the given errors yields instead $Y_{\rm p}$ = 0.2475 $\pm$ 0.0015, the uncertainty 
representing again the error of the mean. This is in line with the helium abundance deduced by the {\it Planck} satellite determination of 
the baryon density, Y$_{\rm p}$ = 0.2471 $\pm$ 0.0003 (e.g., \cite{203}), being consistent with three kinds of neutrinos and a lifetime for free 
neutrons of the order of 870\,s.

\subsection{Dwarf Galaxies Near and Far}

Beyond helium abundances, compact local galaxies experiencing vigorous star formation are cosmologically relevant also in another important
aspect: They may be similar in many respects to those of compact, also vigorously star-forming galaxies in the early Universe. Samples of 
such local low metallicity dwarfs with luminosities greatly affected by young stars have been collected (e.g., \cite{205,206,207}). A detailed comparison 
of compact galaxies at low redshift as analogues of high redshift star-forming galaxies has been carried out by \cite{208}. 25,000 compact star-forming 
$z$ $<$ 1.0 galaxies from the Sloan Digital Sky Survey (SDSS) have been selected and their properties were compared with those of galaxies at 
$z$ $>$ 1.5, deduced by various authors (e.g., \cite{209,210,211,212}). Those low redshift galaxies with H$\beta$ equivalent widths $\ga$100\,\AA\ and 
being located in the star-forming region of the BPT diagram (Baldwin et al., 1981) show remarkable similarities with their higher redshift 
cousins. This concerns fractional oxygen abundances (12 + log(O/H)), stellar masses ($M_*$), far UV absolute magnitudes ($M_{\rm UV}$), star 
formation rates (SFR), specific star-forming rates (sSFR), UV continuum slopes ($\beta_{\rm UV}$), [O{\sc iii}]$\lambda$5007/[O{\sc ii}]$\lambda$3727, 
[Ne{\sc iii}]$\lambda$3868/[O{\sc ii}]$\lambda$3727 and [O/Fe] ratios as well as [O{\sc ii}]$\lambda$3727, [O{\sc iii}]$\lambda$5007, and 
H$\alpha$ equivalent widths. As a consequence, the nearby galaxies can be taken as useful proxies for their more distant counterparts. In 
addition, the nearby galaxies can be traced to absolute magnitudes and linear resolutions that are not yet accessible at high redshifts. 
Thus they provide information that can be used to guide future observations as for example sensitive deep field measurements that may be 
obtained with the {\it James Webb Space Telescope (JWST)} covering galaxies in the early Universe (e.g., \cite{213}).

\subsection{The Role of Dwarfs at Reionization}

Quasars are known to inject copious amounts of Lyman continuum (LyC) photons into the intergalactic medium (IGM), contributing to the 
reionization of the Universe. However, at redshifts $z$ = 6--10, QSOs are rare and thus not numerous enough to drastically change their 
large scale environment alone. While binary stars and less active galactic nuclei may provide additional energetic photons, models indicate
that the participation of a multitude of small low-mass, low-metallicity systems may predominantly contribute the required LyC photons 
for the rapid reionization of the IGM (e.g., \cite{213,214,215}). So far this galaxy population, characterized by low absolute magnitudes and 
luminosities, is below the detection threshold of even the most sensitive telescopes at the required redshifts (\mbox{e.g., \cite{5}}). Furthermore, 
it is difficult to assess the Lyman continuum at the relevant redshifts because of attenuation by the still mostly neutral IGM and interlopers 
at $z$ $<$ 6. 

Therefore, by observing $z$ $>$ 6 galaxies, it is not possible to establish a relationship between the leakage of LyC photons into the 
IGM and the detailed properties of these galaxies. This can only be revealed by observations of galaxies at lower redshift and numerous 
studies have been published to achieve this goal (e.g., \cite{216,217,218,219,220,221,222}). Several ways have been suggested to find galaxies 
with high LyC leakages. Galaxies with a deficit of emission of optical lines representing a low degree of ionization, e.g., [S{\sc ii}], can 
be used in this sense. Other tracers are strong Ly$\alpha$ emission or high [O{\sc iii}]$\lambda$5007/[O{\sc ii}]$\lambda$3727 line ratios 
(see \cite{223} for a particularly extreme galaxy in the latter sense). However, these correlations show a significant amount of scatter. 
The He\,I $\lambda$3889\,\AA\ line in connection with H$\beta$, the [C{\sc iii}]$\lambda$1909 transition, the Mg\,{\sc ii} 2796/2803\,\AA\ 
doublet and the UV H{\sc i} Lyman series were also reported to be suitable tracers to gain insight into the opacity of the sources and 
their ability to allow LyC photons to escape \cite{224,225,226,227,228}. 

Obtaining with the {\it HST} the near-IR SEDs (in the framework of observed wavelengths) from a large number of individual stars, estimating 
their ionizing flux and comparing these fluxes with the amount absorbed by dust and by ionizing the surrounding medium, escape fractions of 
several 10\% were obtained for the nearby dwarf galaxy NGC\,4214 \cite{229}. Another promising way to identify LyC leakers appears to be the analysis 
of Ly$\alpha$ line profiles (e.g., \cite{218,219,230}). Ly$\alpha$ lines peaking slightly off the systemic velocity and, even better, double peaked profiles 
separated by $\la$300\,km\,s$^{-1}$ and down to $\approx$150\,km\,s$^{-1}$ around the systemic velocity in the most suitable cases are indicators of 
significant LyC leaking fractions. The double peaked lines are due to resonance scattering effects with small velocity spacings indicating 
low column densities (e.g., \cite{231}). At high redshifts, the blue peak should be absorbed by neutral foreground gas due to the Hubble expansion, 
unless the source possesses a high peculiar motion away from us and/or a large ionized bubble surrounding it. More locally, due to the 
widespread ionized IGM, this is clearly a lesser problem. Interestingly. a first potential LyC leaker from the reionization epoch, at 
redshift $z$ = 6.593, may have been directly observed by \cite{232}, suggesting the presence of an extended ionized bubble even at such a high 
redshift.

Analyzing the past of the ultra-faint galaxy (UFD) population of the Local Group, the small volume where such galaxies can be detected 
with highest sensitivity,  ref. \cite{233} suggests that the ancestors of such UFDs produced a significant amount (10--50\%) of the UV flux required 
to reionize the IGM. Therefore the Ly$\alpha$ leakage of such or slightly larger galaxies is an essential parameter when considering the 
evolution of the Universe during those early days. This implies that more detailed measurements of the Ly$\alpha$ leakage of nearby metal-poor 
dwarfs are essential to observationally constrain the models and to create a more realistic picture of the time a few hundred million years 
after the Big Bang.

\section{Future Prospects} \label{S7}

Studies of the interstellar medium of dwarf galaxies are presently undergoing a major revolution. Using focal plane array technology, Apertif 
(APERture Tile in Focus) in the Netherlands, ASKAP (Australian SKA Pathfinder) in Australia and MeerKAT in South Africa are designed to cover
simultaneously vast areas of the sky. The latter two of these facilities are also pathfinders for the Square Kilometer Array (SKA), which 
will provide yet another step ahead in sensitivity and angular resolution with respect to low frequency radio waves, covering the $\lambda$ 
$\approx$ 21\,cm line of H{\sc i} and the radio continuum. Furthermore, at optical wavelengths and in the near and mid infrared, the {\it James 
Webb Space Telescope (JWST)} may provide observations with unprecedented quality in the range \mbox{0.5 $\upmu$m--28 $\upmu$m}. Finally, with 
the new generation of large telescopes for optical, near infrared and UV mesurements (e.g., the {\it Extremely Large Telescope (ELT)}) it 
will be possible to determine $L_{\rm UV}$ and $L$(H$\alpha$) of high redshift galaxies, thus studying star formation histories in dwarfs 
outside the local volume. 

With respect to the interstellar medium of dwarf galaxies, many questions are still not settled. What is for example triggering the enhanced 
activity of BCDs and, even more so, the Green Peas? Can the tentatively deduced presence of massive but presumably not supermassive ($M$ 
$<$ 10$^{6}$\,M$_{\odot}$) nuclear engines be verified? What is their number and what are the properties of their parent galaxies with 
respect to other relevant physical paremeters? Will this shed light onto black hole seeds in the distant early Universe? And what will become
the most convincing tool to detect such nuclei: The Ly$\alpha$ line shape, coronal lines or yet another tracer? How frequent are galaxies consisting
almost entirely of neutral hydrogen gas as a function of redshift? Are they much more common at higher redshifts than locally? And is there 
evidence for high-$J$ CO lines in clouds with nascent super star clusters complemented by hot dust and perhaps even by an X-ray dominated region
(XDR) caused by the X-ray emission of the numerous newly formed massive stars?  These and many other questions, some of them addressed in the 
accompanying articles, will keep dwarf galaxies in the focus of contemporary astrophyical research and will continue to attract a high degree 
of interest in these seemingly inconspicious but very numerous galaxies, that provide a great amount of information on the evolution of the 
Universe, at present times, in the distant past and as a consequence also on its future.

\vspace{6pt}

\acknowledgments{We appreciate the very positive and constructive comments of the anonymous referees. 
Y.I.I. acknowledges support from the National Academy of Sciences of Ukraine (Project 0121U109612: 
`Dynamics of particles and collective excitations in high energy physics, astrophsyics and quantum 
microsystems'). The article made use of the Astrophysics Data System and the NASA/IPAC Extragalactic 
Database.}

\reftitle{References}




\end{document}